\documentclass[12pt]{article}
\usepackage{graphicx}
\usepackage{amsmath}
\usepackage{amsthm}
\usepackage{bbm}
\usepackage{amsfonts}
\usepackage{enumerate}
\usepackage{subfigure}
\usepackage{natbib}
\usepackage{threeparttable}
\usepackage[geometry]{ifsym}
\usepackage[nolists]{endfloat}
\usepackage{color}
\usepackage{url}

\newcommand{\bs}{\boldsymbol}

\newtheorem{thm}{Theorem}
\newtheorem{lem}{Lemma}

\newtheorem{cor}{Corollary}

\newtheorem{prop}{Proposition}
\newtheorem{assum}{Assumption}
\newcommand{\tu}{\tilde{U}}
\newcommand{\ta}{\tilde{A}}

\def\urltilda{\kern -.15em\lower .7ex\hbox{\~{}}\kern .04em}
\def\urldot{\kern -.10em.\kern -.10em}
\def\urlhttp{http\kern -.10em\lower -.1ex\hbox{:}\kern -.12em\lower 0ex\hbox{/}\kern -.18em\lower 0ex\hbox{/}}
\def\myhyph{-\penalty0\hskip0pt\relax}

\begin{document}

\title{Locally Adaptive Bayes Nonparametric Regression via \\ Nested Gaussian Processes
\author{Bin Zhu and David B. Dunson$^*$}
}

\date{}
\maketitle
\let\thefootnote\relax\footnotetext{$^*$Bin Zhu is Postdoctoral Associate, Department of Statistical Science and Center for Human Genetics, 
Duke University, Durham, NC 27708, (Email: \emph{bin.zhu@duke.edu}). David B. Dunson is Professor, Department of Statistical Science, Duke University, Durham, NC 27708, (Email: \emph{dunson@stat.duke.edu}). }

\centerline{\textbf{Abstract}}
\noindent 
We propose a nested Gaussian process (nGP) as a locally adaptive prior for Bayesian nonparametric regression. Specified through a set of stochastic differential equations (SDEs), the nGP imposes a Gaussian process prior for the function's $m$th-order derivative.  The nesting comes in through including a local instantaneous mean function, which is drawn from another Gaussian process inducing adaptivity to locally-varying smoothness.  We discuss the support of the nGP prior in terms of the closure of a reproducing kernel Hilbert space, 
and consider theoretical properties of the posterior.  The posterior mean under the nGP prior is shown to be equivalent to the minimizer of a nested penalized sum-of-squares involving penalties for both the global and local roughness of the function.  Using highly-efficient Markov chain Monte Carlo for posterior inference, the proposed method performs well in simulation studies compared to several alternatives, and is scalable to massive data, illustrated through a proteomics application.\\
\noindent \textbf {Key words}: Bayesian nonparametric regression; Nested Gaussian processes; Nested smoothing spline; Penalized sum-of-square; Reproducing kernel Hilbert space; Stochastic differential equations. 

\newpage
\section{Introduction}
\label{sec:intro}
\begin{figure}
\centering
\subfigure[]{\includegraphics[width=0.48\textwidth]{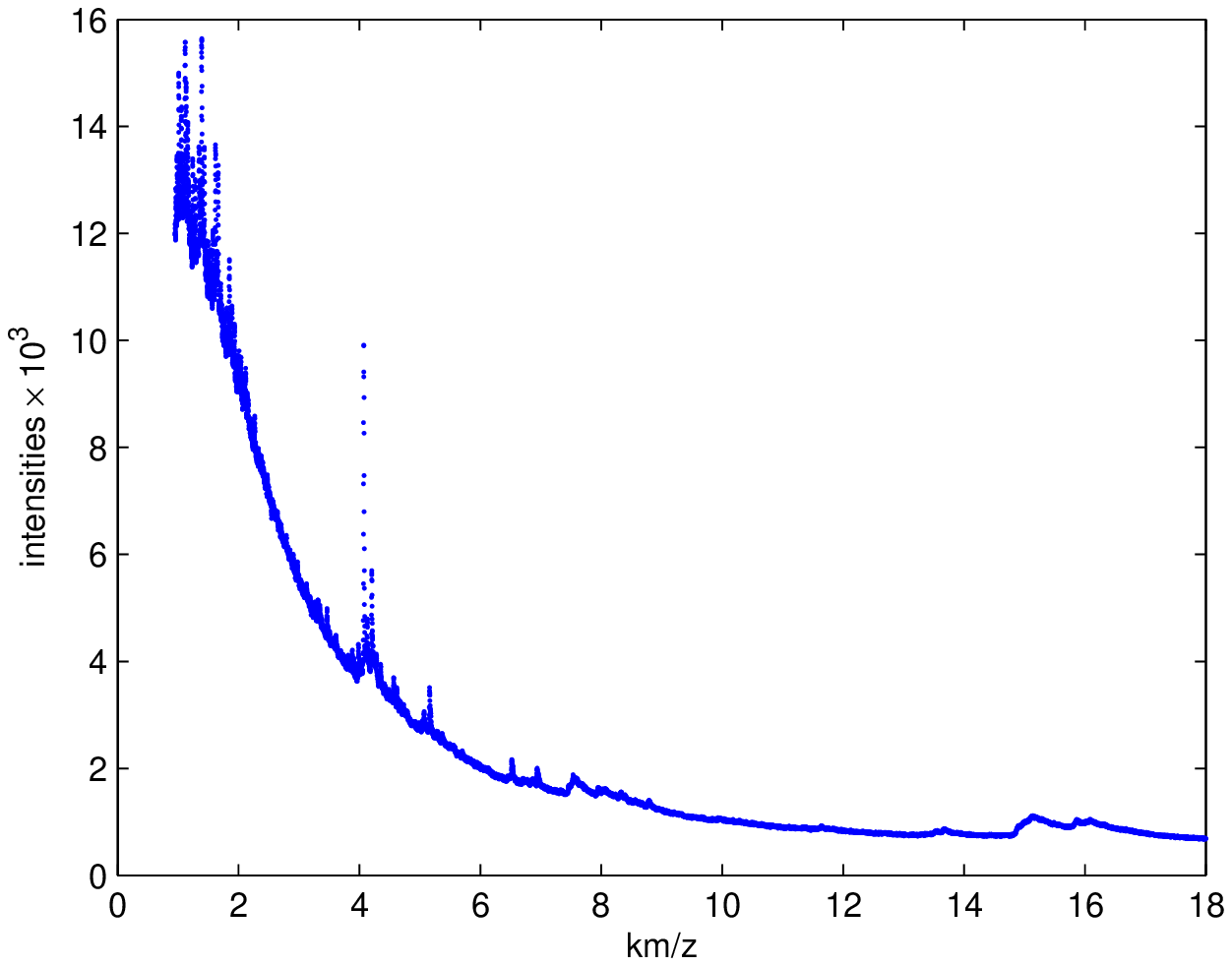}}
\subfigure[]{\includegraphics[width=0.48\textwidth]{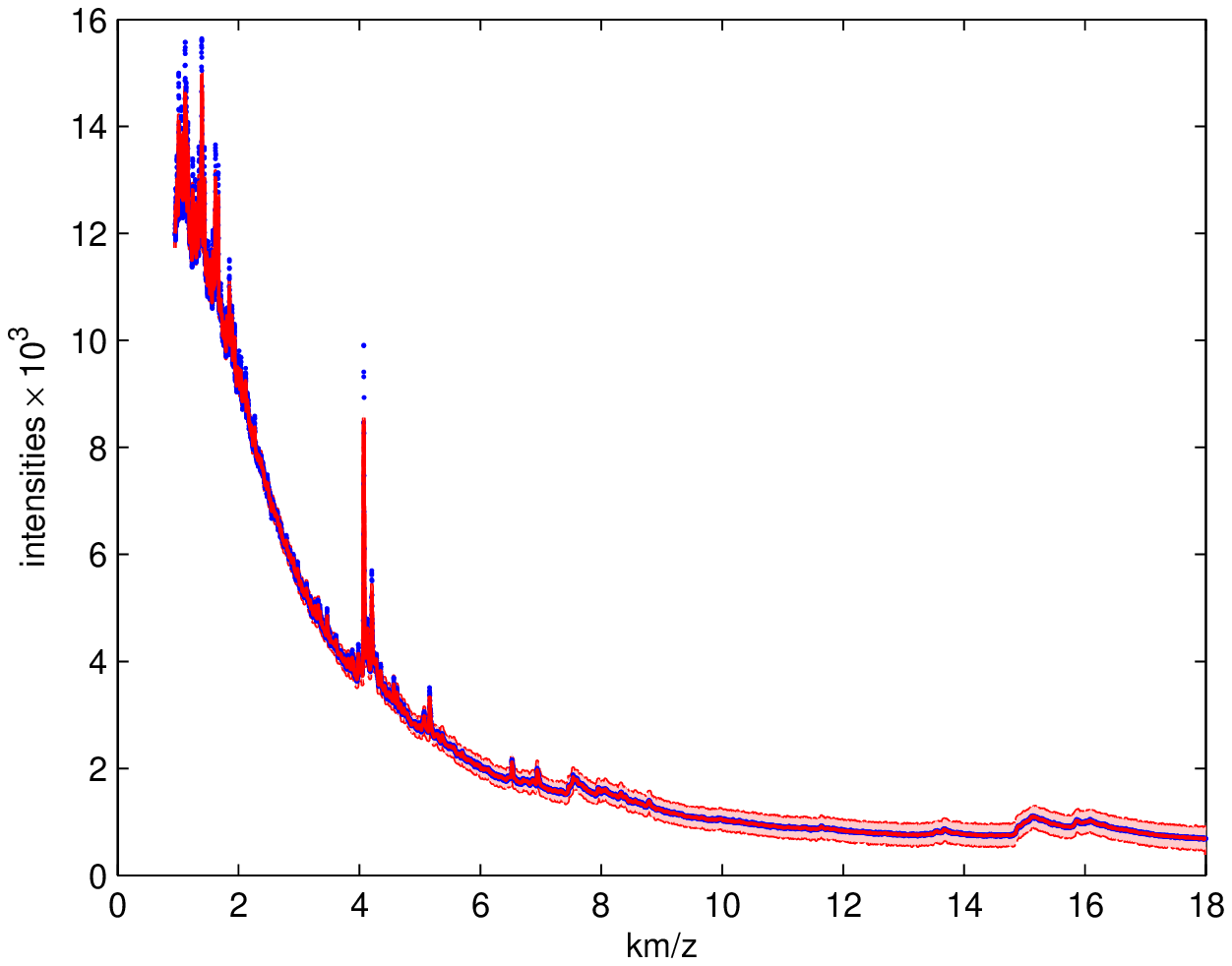}}\\
\subfigure[]{\includegraphics[width=0.48\textwidth]{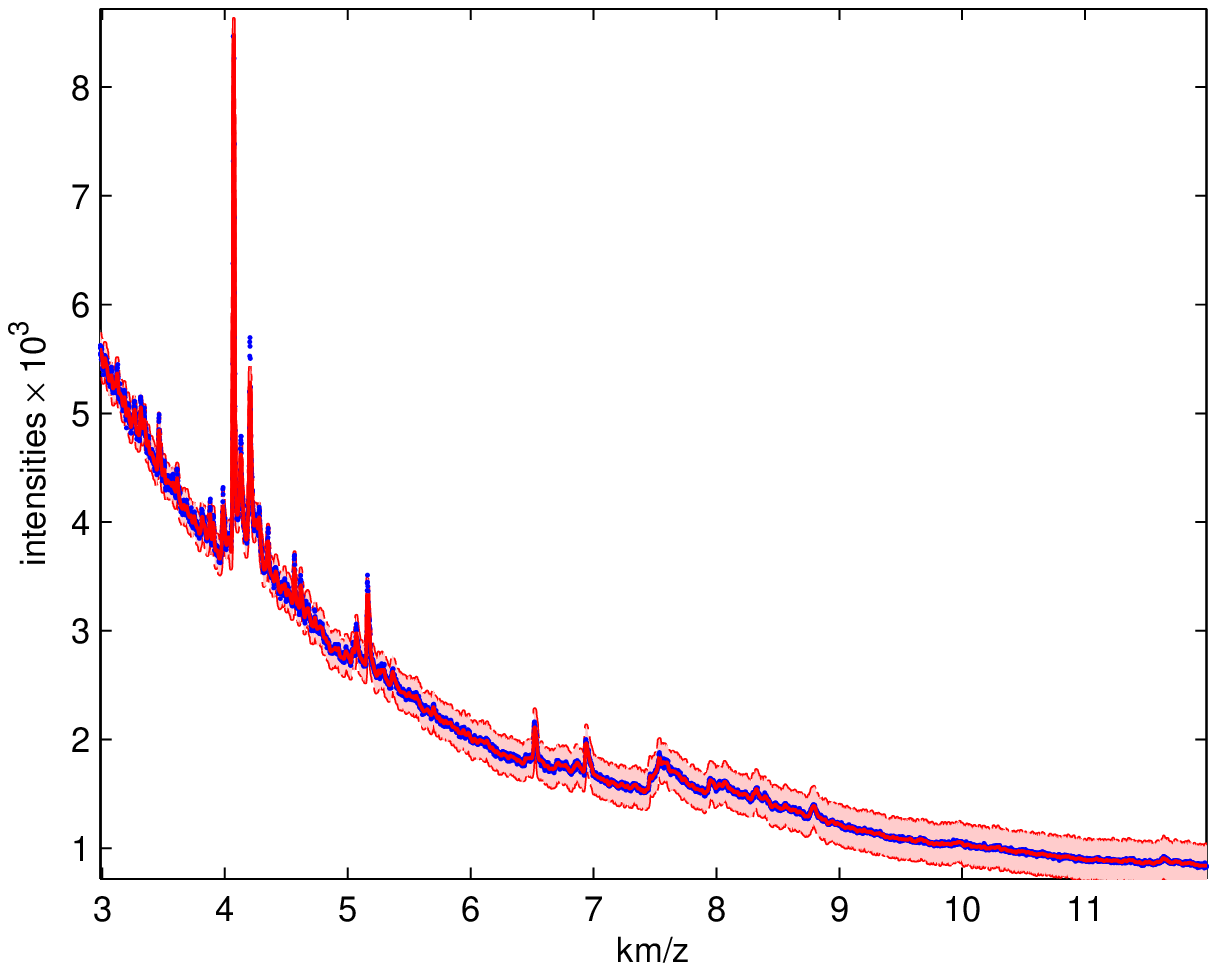}}
\subfigure[]{\includegraphics[width=0.495\textwidth]{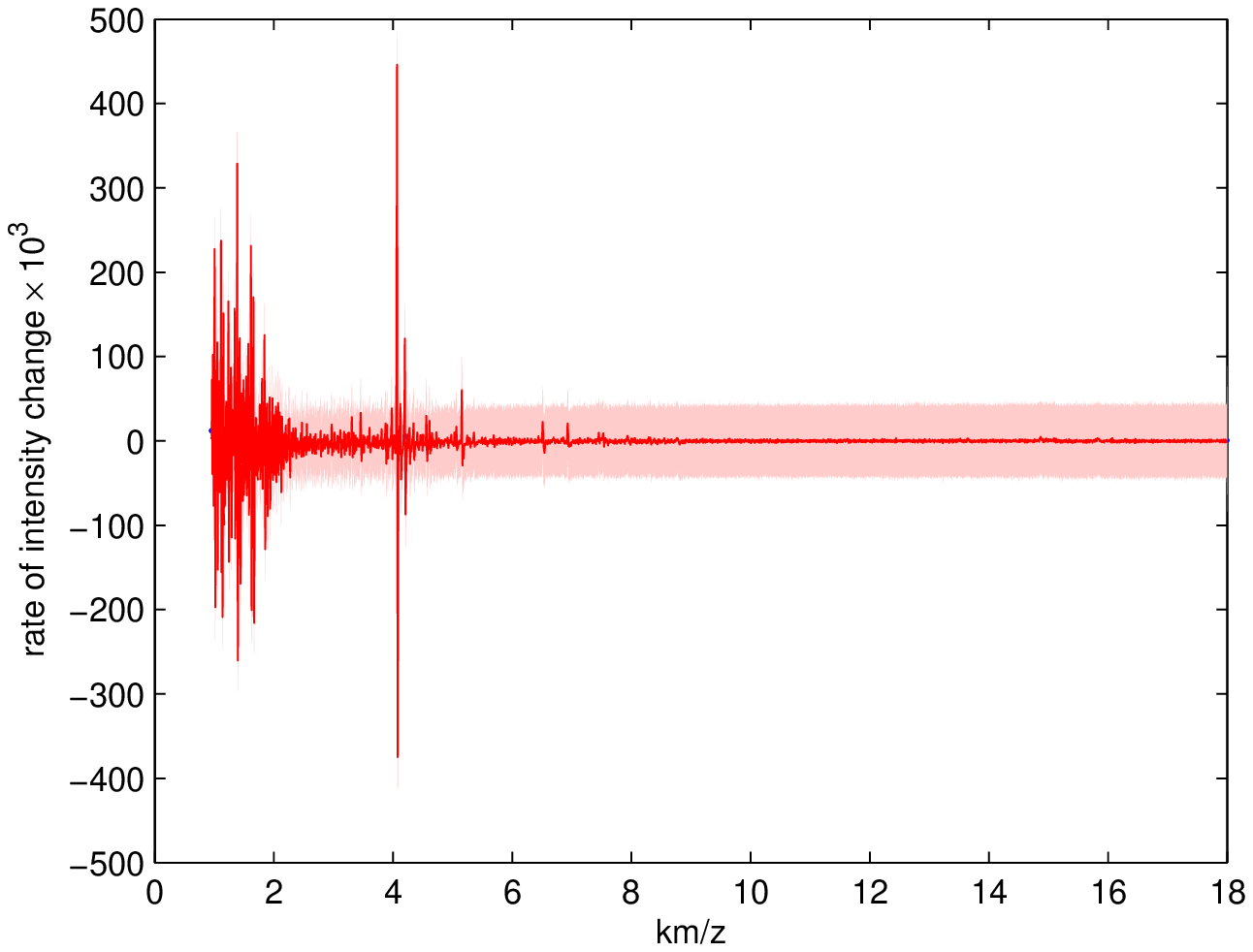}}
\caption{\label{fig:MCMC} 
(a) Plot of protein mass spectrometry data:  observed intensities versus mass to charge ratio m/z; 
(b) Posterior mean ({\color{red} \textemdash}) and $95\%$ credible interval of $U$ (red shades); 
(c) Posterior mean and $95\%$ credible interval of $U$ for a local region; 
(d) Posterior mean and $95\%$ credible interval of rate of intensity changes $DU$.}
\end{figure}

We consider the nonparametric regression problem
\begin{equation}
\label{eq:obs}
Y(t) = U(t) + \varepsilon(t),\;\;t \in \mathcal{T} = [t_0,t_U],
\end{equation}
where $U: \mathcal{T} \to \mathbb{R}\;$ is an unknown mean regression function to be estimated at $\mathcal{T}_o = \{ t_0, t_1, t_2, \ldots, t_J < t_U \}$, $t_0=0$, and $\bs{\varepsilon}=\left[\varepsilon(t_1),\varepsilon(t_2),\cdots, \varepsilon(t_J)\right]^\prime \sim \mathsf{N}_J(\bs{0},  \sigma^{2}_{\varepsilon} \bs{I})$ a $J$-dimensional multivariate normal distribution with mean vector $\bs{0}$ and covariance matrix $\sigma^{2}_{\varepsilon} \bs{I}$. We are particularly interested in allowing the smoothness of $U$ to vary locally as a function of $t$.  For example, consider the protein mass spectrometry data in panel (a) of Figure \ref{fig:MCMC}.  There are clearly regions of $t$ across which the function is very smooth and other regions in which there are distinct spikes, with these spikes being quite important. An additional challenge is that the data are generated in a high-throughput experiment with $J=11,186$ observations.  Hence, we need a statistical model which allows locally-varying smoothness, while also permitting efficient computation even when data are available at a large number of locations along the function.

A commonly used approach for nonparametric regression is to place a Gaussian process (GP) prior \citep{neal1998regression,rasmussen2006gaussian,shi1011gaussian} on the unknown $U$, where the GP is usually specified by its mean and covariance function (e.g. squared exponential). The posterior distribution of $U(\mathcal{T}_o)$ can be conveniently obtained as a multivariate Gaussian distribution.  When carefully-chosen hyperpriors are placed on the parameters in the covariance kernel, GP priors have been shown to lead to large support, posterior consistency \citep{ghosal2006posterior,choi2007posterior} and even near minimax optimal adaptive rates of posterior contraction \citep{van2008rates}.  However, the focus of this literature has been on isotropic Gaussian processes, which have a single bandwidth parameter controlling global smoothness, with the contraction rate theory assuming the true function has a single smoothness level.  There has been applied work allowing the smoothness of a multivariate regression surface to vary in different directions by using predictor-specific bandwidths in a GP with a squared exponential covariance \citep{savitsky2011variable,zou2010nonparametric}.  \citet*{Bhattacharya2011} recently showed that a carefully-scaled anisotropic GP leads to minimax optimal adaptive rates in anisotropic function classes including  when the true function depends on a subset of the predictors.  However, the focus was on allowing a single smoothness level for each predictor, while our current interest is allowing smoothness to vary locally in nonparametric regression in a single predictor.

There is a rich literature on locally-varying smoothing.  One popular approach relies on free knot splines, for which various strategies have been proposed to select the number of knots and their locations, including stepwise forward and/or backward knots selection \citep{friedman1989flexible,friedman1991multivariate,luo1997hybrid}, accurate knots selection scheme \citep{zhou2001spatially} and Bayesian knots selection \citep{smith1996nonparametric, denison1998automatic,dimatteo2001bayesian} via Gibbs sampling \citep{george1993variable} or reversible jump Markov chain Monte Carlo \citep{green1995reversible}. Although many of these methods perform well in simulations, such free knot approaches tend to be highly computationally demanding making their implementation in massive data sets problematic.

In addition to free knot methods, adaptive penalization approaches have also been proposed.  An estimate of $U$ is obtained as the minimizer of a penalized sum of squares including a roughness penalty with a spatially-varying smoothness parameter \citep{Wahba1995adpative, ruppert2000spatially, pintore2006spatially,crainiceanu2007spatially}. Other smoothness adaptive methods include wavelet shrinkage \citep{donoho1995adapting}, local polynomial fitting with variable bandwidth \citep{fan1995data}, L-spline \citep{abramovich1996improved,heckman2000penalized}, mixture of splines \citep{wood2002bayesian} and linear combination of kernels with varying bandwidths \citep{Wolpert2011Stochastic}. The common theme of these approaches is to reduce the constraint on the single smoothness level assumption and to implicitly allow the derivatives of $U$, a common measurement of the smoothness of $U$, to vary over $t$. 

In this paper, we instead propose a nested Gaussian process (nGP) prior to explicitly model the expectation of the derivative of $U$ as a function of $t$ and to make full Bayesian inference using an efficient Markov chain Monte Carlo (MCMC) algorithm scalable to massive data. More formally, our nGP prior specifies a GP for $U$'s $m$th-order derivative $D^m U$ centered on a local instantaneous mean function $A: \mathcal{T} \to \mathbb{R}\;$ which is in turn drawn from another GP. Both GPs are defined by stochastic differential equations (SDEs), related to the method proposed by \citet{zhu2011stochastic}. However, \citet{zhu2011stochastic} centered their process on a parametric model, while we instead center on a higher-level GP to allow nonparametric locally-adaptive smoothing.  Along with the observation equation \eqref{eq:obs}, SDEs can be reformulated as a state space model \citep{durbin2001time}. This reformulation facilitates the application of simulation smoother \citep{durbin2002simple},  an efficient MCMC algorithm with $\mathcal{O}(J)$ computational complexity which is essential to deal with large scale data. We will show that the nGP prior has large support and its posterior distribution is asymptotically consistent.  In addition, the posterior mean or 
mode of U under the nGP prior can be shown to correspond to the minimizer of a penalized sum of squares with nested penalty functions.

The remainder of the paper is organized as follows. Section \ref{sec:nGP} defines the nGP prior and discusses some of its properties.  Section \ref{sec:comp} outlines an efficient Markov chain Monte Carlo (MCMC) algorithm for posterior computation. Section \ref{sec:simu} presents simulation studies.
The proposed method is applied to a mass spectra dataset in Section \ref{sec:app}. Finally, Section \ref{sec:dis} contains several concluding remarks and outlines some future directions. 
 
\section{Nested Gaussian Process Prior}
\label{sec:nGP}
\subsection{Definition and Properties}
The nGP defines a GP prior for the mean regression function $U$ and the local instantaneous mean function $A$ through the following SDEs with parameters $\sigma_{U}\in \mathbb{R}^+ $ and $\sigma_{A}\in \mathbb{R}^+ $:
\begin{align}
\label{eq:sde_u}
D^{m} U(t) &= A(t) + \sigma_{U} \dot{W}_U(t),\quad \; m \in \mathbb{N}\geq 2
\\
\label{eq:sde_a}
D^{n} A(t) &= \sigma_{A} \dot{W}_A(t),\quad \; n \in \mathbb{N}\geq 1
\end{align} 
where $\dot{W}_U(t)$ and $\dot{W}_A(t)$ are two independent Gaussian white noise processes with mean function $\textsf{E}\{\dot{W}_U(t)\}=\textsf{E}\{\dot{W}_A(t)\}=0$ and covariance function $\textsf{E}\{\dot{W}_U(t)\dot{W}_U(t^\prime)\}=\textsf{E}\{\dot{W}_A(t)\dot{W}_A(t^\prime)\}=\delta(t-t^\prime)$ a delta function. The initial value of $U$ and its derivatives up to order $m-1$ at $t_0=0$ are denoted as $\bs{\mu}=( 
\mu_{0}, \mu_{1},\cdots,\mu_{m-1})^\prime \sim \textsf{N}_m(\bs{0}, \sigma^2_{\mu}\bs{I})$. Similarly, the initial values of $A$ and its derivatives till order $n-1$ at $t_0=0$ are denoted as $\bs{\alpha}=( \alpha_{0}, \alpha_{1},\cdots, \alpha_{n-1})^\prime \sim \textsf{N}_n(\bs{0}, \sigma^2_{\alpha}\bs{I})$. In addition, we assume that $\bs{\mu}$, $\bs{\alpha}$, $\dot{W}_U(\cdot)$ and $\dot{W}_A(\cdot)$ are mutually independent. 
The definition of nGP naturally induces a prior for $U$ with varying smoothness. Indeed, the SDE \eqref{eq:sde_u} suggests that $\textsf{E}\{ D^m U(t) \mid A(t)\}=A(t)$. Thus, the smoothness of U,  measured by $D^m U$,  is expected to be centered on a function $A$ varying over $t$. 


We first recall the definition of the reproducing kernel Hilbert space (RKHS) generated by the zero-mean Gaussian process $W=\{W(t): t\in \mathcal{T}\}$ and the results on the support of $W$, which will be useful to explore the theoretical properties of the nGP prior.   
Let $(\Omega,\mathcal{A}, P)$ be the probability space for $W$ such that for any $t_1,t_2,\dots,t_k \in \mathcal{T}$ with $k \in \mathbb{N}\,$, $\{W(t_1),W(t_2),\dots,W(t_k)\}^\prime$ follow a zero-mean multivariate normal distribution with covariance matrix induced through the covariance function $\mathcal{K}_W: \mathcal{T}\times\mathcal{T} \to \mathbb{R}\,$, defined by $\mathcal{K}_W(s,t)=E\{W(s)W(t)\}$. The RKHS $\mathcal{H}_{\mathcal{K}_W}$ generated by $W$ is the completion of the linear space of all functions   
\begin{equation*}
t \mapsto \sum_{i=1}^k a_i \mathcal{K}_W(s_i, t),\quad a_1,\ldots,a_k \in \mathbb{R}\,, s_1,\ldots, s_k \in \mathcal{T}, k \in \mathbb{N}\;,
\end{equation*}
with the inner product
\begin{equation*}
\left\langle \, \sum_{i=1}^k a_i\mathcal{K}_W(s_i,\cdot) , \sum_{j=1}^l b_j\mathcal{K}_W(t_j,\cdot) \right\rangle_{\mathcal{H}_{\mathcal{K}_W}}=\;\;\sum_{i=1}^k\sum_{j=1}^la_ib_j\mathcal{K}_W(s_i,t_j),
\end{equation*}
which satisfies the reproducing property $f(t) = \left\langle \, f ,  \mathcal{K}_W(t,\cdot) \right\rangle_{\mathcal{H}_{\mathcal{K}_W}}$ for any $f \in \mathcal{H}_{\mathcal{K}_W}:\mathcal{T} \to \mathbb{R}\,$.

With the specification of the RKHS $\mathcal{H}_{\mathcal{K}_W}$, we are able to define the support of W as the closure of $\mathcal{H}_{\mathcal{K}_W}$\citep[Lemma 5.1, ][]{van2008reproducing}. We apply this definition to characterize the support of the nGP prior,  which is formally stated in Theorem \ref{thm:RKHS_nGP}. Its proof requires the results of the following lemma.
\begin{lem}
\label{lem:GPs_nGP}
The nested Gaussian process $U$ can be written as $U(t)=\tilde{U}_0(t)+\tilde{U}_1(t)+\tilde{A}_0(t)+\tilde{A}_1(t)$, the summation of mutually independent Gaussian processes with the corresponding mean functions $\textsf{E}\left\{{U}_0(t)\right\}=\textsf{E}\left\{{U}_1(t)\right\}=
 \textsf{E}\left\{{A}_0(t)\right\}=\textsf{E}\left\{{A}_1(t)\right\}=0$ and covariance functions 
\begin{align*}
\mathcal{K}_{\tilde{U}_0}(s,t)&=\sigma^2_{\mu}\mathcal{R}_{\tilde{U}_0}(s,t)=\sigma^2_{\mu}\sum_{i=0}^{m-1}\phi_i(s)\phi_i(t),\\
\mathcal{K}_{\tilde{U}_1}(s,t)&=\sigma^2_U\mathcal{R}_{\tilde{U}_1}(s,t)=\sigma^2_U\int_{\mathcal{T}}G_m(s,u)G_m(t,u)du,\\
\mathcal{K}_{\tilde{A}_0}(s,t)&=\sigma^2_{\alpha}\mathcal{R}_{\tilde{A}_0}(s,t)=\sigma^2_{\alpha}\sum_{i=0}^{n-1}\phi_{m+i}(s)\phi_{m+i}(t),\\
\mathcal{K}_{\tilde{A}_1}(s,t)&=\sigma^2_A\mathcal{R}_{\tilde{A}_1}(s,t)=\sigma^2_A\int_{\mathcal{T}}G_{m+n}(s,u)G_{m+n}(t,u)du,
\end{align*}
respectively, where $\phi_i(t)=\frac{t^i}{i!}$ and $G_m(s,u)=\frac{(s-u)_{+}^{m-1}}{(m-1)!}$.
\end{lem}
\noindent The proof is in Appendix \ref{sec:appendix_a}. 
\begin{thm}
\label{thm:RKHS_nGP}
The support of nested Gaussian process U is the closure of RKHS $\mathcal{H}_{\mathcal{K}_U}=\mathcal{H}_{\mathcal{K}_{\tilde{U}_0}}\oplus\mathcal{H}_{\mathcal{K}_{\tilde{U}_1}}\oplus
\mathcal{H}_{\mathcal{K}_{\tilde{A}_0}}\oplus\mathcal{H}_{\mathcal{K}_{\tilde{A}_1}}$, the direct sum of RKHSs $\mathcal{H}_{\mathcal{K}_{\tilde{U}_0}}$, $\mathcal{H}_{\mathcal{K}_{\tilde{U}_1}}$,
$\mathcal{H}_{\mathcal{K}_{\tilde{A}_0}}$ and  $\mathcal{H}_{\mathcal{K}_{\tilde{A}_1}}$ with reproducing kernels $\mathcal{K}_{\tilde{U}_0}(s,t)$, $\mathcal{K}_{\tilde{U}_1}(s,t)$, $\mathcal{K}_{\tilde{A}_0}(s,t)$ and $\mathcal{K}_{\tilde{A}_1}(s,t)$ respectively. 
\end{thm}
\noindent The proof is in Appendix \ref{sec:appendix_a}. By Corollary \ref{cor:GP_pss}, it is of interest to note that $\mathcal{H}_{\mathcal{K}_U}$ includes a subspace $\mathcal{H}_{\mathcal{K}_{\tilde{U}}}$, which is the RKHS for the polynomial smoothing spline \citep[][Section 1.5]{wahba1990spline}.

\begin{cor}
\label{cor:GP_pss}
The support of the Gaussian process $\tilde{U}=\tilde{U}_0+\tilde{U}_1$ as the prior for polynomial smoothing spline is the closure of RKHS $\mathcal{H}_{\mathcal{K}_{\tilde{U}}}=\mathcal{H}_{\mathcal{K}_{\tilde{U}_0}}\oplus\mathcal{H}_{\mathcal{K}_{\tilde{U}_1}}$ with $\mathcal{H}_{\mathcal{K}_{\tilde{U}}} \subset \mathcal{H}_{\mathcal{K}_{U}}$. 
\end{cor}
\noindent The proof is in Appendix \ref{sec:appendix_a}. Hence, it is obvious that the nGP prior includes GP prior for polynomial smoothing spline as a special case when $\sigma_\alpha^2 \to 0$ and $\sigma_A^2 \to 0$. 

The nGP prior can generate functions $U$ arbitrarily close to any function $U_0$ in the support of the prior.  From Theorem \ref{thm:RKHS_nGP} it is clear that the support is large and hence the sample paths from the proposed prior can approximate any function in a broad class.  As a stronger property, it is also appealing that the posterior distribution concentrate in arbitrarily small neighborhoods of the true function $U_0$ which generated the data as the sample size $J$ increases, with this property referred to as posterior consistency. 
More formally, a prior $\Pi$ on $\Theta$ achieves posterior consistency at the true parameter $\theta_0$ if for any neighborhoods $\mathcal{U}_{\epsilon}$, the posterior distribution $\Pi\left(\mathcal{U}_{\epsilon} \mid Y_1, Y_2, \dots, Y_J\right) \to 1$ almost surely under $\Pi_{\theta_0}$,  the true joint distribution of observations $\{Y_j\}_{j=1}^J$. For our case, the parameters $\theta=(U,\sigma_\varepsilon)$ lie in the product space $\Theta=\mathcal{H}_{\mathcal{K}_U}\times\mathbb{R}^+$ and have a prior $\Pi_{\theta}=\Pi_{U}\times\Pi_{\sigma_\varepsilon}$, for which $\Pi_{U}$ is an nGP prior for $U$ and $\Pi_{\sigma_\varepsilon}$ is a prior distribution for $\sigma_\varepsilon$.
The $L_1$ neighborhood of $\theta_0=(U_0,\sigma_{\varepsilon,0})$ is defined as $\mathcal{U}_{\epsilon}=\left\{ (U,\sigma_\varepsilon)\ : || U-U_0 ||_1 = \int_0^{t_U} |U(t)-U_0(t)|dt < \epsilon, \left |{\sigma_\varepsilon}-{\sigma_{\varepsilon,0}}\right|<\epsilon \right\}$.  
We further specify a couple of regularity conditions given by:
\begin{assum} \label{assum:design}
$t_j
$ arises according to an infill design: for each $S_j=t_{j+1}-t_j$, there exists a constant $0<C_d\leq 1$ such that $\max_{1 \leq j < J} S_j < \frac{t_U}{C_dJ}$.
\end{assum}
\begin{assum}\label{assum:exp}
The prior distributions $\Pi_{\sigma^{2}_{\mu}}$, $\Pi_{\sigma^{2}_{U}}$, $\Pi_{\sigma^{2}_{\alpha}}$ and $\Pi_{\sigma^{2}_{A}}$ satisfy an exponential tail condition. Specifically, there exist sequences $M_J$, $\sigma^2_{\mu,J}$, $\sigma^2_{U,J}$, $\sigma^2_{\alpha,J}$ and $\sigma^2_{A,J}$ such that: (i) $\Pi_{\sigma^{2}_{\mu}}(\sigma^{2}_{\mu}>\sigma^{2}_{\mu,J})=e^{-C_\mu J}$, $\Pi_{\sigma^{2}_{U}}(\sigma^{2}_{U}>\sigma^{2}_{U,J})=e^{-C_UJ}$,
$\Pi_{\sigma^{2}_{\alpha}}(\sigma^{2}_{\alpha}>\sigma^{2}_{\alpha,J})=e^{-C_\alpha J}$ and $\Pi_{\sigma^{2}_{A}}(\sigma^{2}_{A}>\sigma^{2}_{A,J})=e^{-C_A J}$, for some positive constants $C_\mu$, $C_U$, $C_\alpha$ and $C_A$; (ii) $M_J^2\sigma^{-2}_J \geq C_gJ$,  for every $C_g > 0$ and $\sigma^{-2}_J$,  the minimal element of $\{\sigma^{-2}_{\mu,J}, \sigma^{-2}_{U,J}, \sigma^{-2}_{\alpha,J}, \sigma^{-2}_{A,J} \}$. 
\end{assum} 
\begin{assum}\label{assum:posprior}
The prior distribution $\Pi_{\sigma_{\varepsilon}}$ is continuous and the $\sigma_{\varepsilon, 0}$ lies in the support of $\Pi_{\sigma_{\varepsilon}}$ 
\end{assum}
\noindent Under those specifications and regularity conditions, the results on strong posterior consistency for the Bayes nonparametric regression with nGP prior is given as follows.      
\begin{thm}
\label{thm:pos_con} Let $\{Y_j\}_{j=1}^J$ be the independent but non-identical observations following normal distributions $\{\textsf{N}_1(U(t_j), \sigma^2_\varepsilon)\}_{j=1}^J$ with unknown mean function $U$ and unknown $\sigma^2_\varepsilon$ at design points $t_1,t_2,\dots t_J$. Suppose $U$ follows an nGP prior and the Assumptions \ref{assum:design},\ref{assum:exp} and \ref{assum:posprior} hold. Then for every $\theta_0 \in \Theta$ and  every $\epsilon > 0$,
\begin{equation*}
\Pi\left(\mathcal{U}_{\epsilon} \mid Y_1, Y_2, \dots, Y_J \right) \to 1 \text{  a.s. under } \Pi_{\theta_0}.
\end{equation*}
 
\end{thm}
\noindent The proof is based on the strong consistency theorem by \citet{choi2007posterior} and is detailed in Appendix \ref{sec:appendix_a}.    
\subsection{Connection to Nested Smoothing Spline}
We show in Theorem \ref{thm:nGPequalnSS} that the posterior mean of $U$ under an nGP prior can be related to the minimizer, namely the nested smoothing spline (nSS) $\hat{U}$,  of the following penalized sum-of-squares with nested penalties,
\begin{equation}
\label{eq:npss}
\textsf{nPSS}(t)=
\frac{1}{J}\sum_{j=1}^J\left\{Y(t_j)-U(t_j)\right\}^2 + \lambda_{U}\int_{\mathcal{T}}\left\{D^{m}U(t)-A(t)\right\}^2dt +
\lambda_{A}\int_{\mathcal{T}}\left\{D^{n}A(t)\right\}^2dt,
\end{equation}
where $\lambda_{U}\in \mathbb{R}^+$ and $\lambda_{A}\in \mathbb{R}^+$ are the smoothing parameters which control the smoothness of unknown functions $U(t)$ and $A(t)$ respectively.  The following Theorem \ref{thm:nSS} and Corollary \ref{cor:nSS_coef} provide the explicit forms for nSS, for which the proofs are included in Appendix \ref{sec:appendix_a}. 

\begin{thm}
\label{thm:nSS}
The nested smoothing spline $\hat{U}(t)$ has the form 
\begin{align*}
\hat{U}(t)
&=
\sum_{i=0}^{m-1}\mu_i\phi_i(t)+
\sum_{j=1}^{J}\nu_j\mathcal{R}_{\tilde{U}_1}(t_j,t)+
\sum_{i=0}^{n-1}\alpha_i\phi_{m+i}(t)+
\sum_{j=1}^{J}\beta_j\mathcal{R}_{\tilde{A}_1}(t_j,t)\\ 
&=
\bs{\mu}^\prime\bs{\phi}_\mu(t)+
\bs{\nu}^\prime\bs{R}_{\tu}(t)+
\bs{\alpha}^\prime\bs{\phi}_\alpha(t)+
\bs{\beta}^\prime\bs{R}_{\ta}(t),
\end{align*}
where
$\bs{\mu}=(\mu_0,\mu_1,\cdots,\mu_{m-1})^\prime$, 
$\bs{\nu}=(\nu_1,\nu_1,\cdots,\nu_J)^\prime$, $\bs{\alpha}=(\alpha_0,\alpha_1,\cdots,\alpha_{n-1})^\prime$ and
$\bs{\beta}=(\beta_1,\beta_2,\cdots,\beta_j)^\prime$
are the coefficients for the bases
\begin{align*}
\bs{\phi}_\mu(t)&=\{\phi_0(t),\phi_1(t),\cdots,\phi_{m-1}(t)\}^\prime,\quad
\bs{R}_{\tu}(t)=\{\mathcal{R}_{\tu_1}(t_1,t),\mathcal{R}_{\tu_1}(t_2,t),\cdots,\mathcal{R}_{\tu_1}(t_J,t)\}^\prime,\\
\bs{\phi}_\alpha(t)&=\{\phi_m(t),\phi_{m+1}(t),\cdots,\phi_{m+n-1}(t)\}^\prime,\quad
\bs{R}_{\ta}(t)=\{\mathcal{R}_{\ta_1}(t_1,t),\mathcal{R}_{\ta_1}(t_2,t),\cdots,\mathcal{R}_{\ta_1}(t_J,t)\}^\prime.\\
\end{align*} 
In addition, the nested penalized sum-of-squares can be written as
\begin{align*}
\textsf{nPSS}(t)=
&\frac{1}{J}\left(\bs{Y}-\bs{\phi}_{\mu}\bs{\mu}-\bs{R}_{\tu}\bs{\nu}-\bs{\phi}_{\alpha}\bs{\alpha}-\bs{R}_{\ta}\bs{\beta}\right)^\prime
\left(\bs{Y}-\bs{\phi}_{\mu}\bs{\mu}-\bs{R}_{\tu}\bs{\nu}-\bs{\phi}_{\alpha}\bs{\alpha}-\bs{R}_{\ta}\bs{\beta}\right)\\
&+\lambda_U\bs{\nu}^\prime\bs{R}_{\tu}\bs{\nu}+\lambda_A\bs{\beta}^\prime\bs{R}_{\ta}\bs{\beta},
\end{align*}
where
\vspace{-15pt} 
\begin{align*} 
\bs{Y}&=\{Y(t_1),Y(t_1),\cdots,Y(t_J)\}^\prime,\\
\bs{\phi}_\mu&=\{\bs{\phi}_\mu(t_1),\bs{\phi}_\mu(t_2),\cdots,\bs{\phi}_\mu(t_J)\}^\prime, \quad 
\bs{\phi}_\alpha=\{\bs{\phi}_\alpha(t_1),\bs{\phi}_\alpha(t_2),\cdots,\bs{\phi}_\alpha(t_J)\}^\prime,\\
\bs{R}_{\tu}&=\{\bs{R}_{\tu}(t_1),\bs{R}_{\tu}(t_2),\cdots,\bs{R}_{\tu}(t_J)\},\quad
\bs{R}_{\ta}=\{\bs{R}_{\ta}(t_1),\bs{R}_{\ta}(t_2),\cdots,\bs{R}_{\ta}(t_J)\}.
\end{align*}
\end{thm}

\begin{cor}
\label{cor:nSS_coef}
The coefficients $\bs{\mu}$, $\bs{\nu}$, $\bs{\alpha}$ and $\bs{\beta}$ of the nested smoothing spline $\hat{U}(t)$ in Theorem \ref{thm:nSS}  are given as
\begin{align*}
\bs{\mu} &=
\bs{\Sigma}^{-1}_{\mu \mid \alpha}
\bs{\phi}_{\mu \mid \alpha}
\bs{S}^{-1}
\bs{Y}, \\
\bs{\nu} &= \bs{S}^{-1}
\left\{
\bs{I}
-
\left(
\bs{\phi}_{\mu}\bs{\Sigma}^{-1}_{\mu \mid \alpha}\bs{\phi}_{\mu \mid \alpha}
+
\bs{\phi}_{\alpha}\bs{\Sigma}^{-1}_{\alpha \mid \mu}\bs{\phi}_{\alpha \mid \mu}
\right)
\bs{S}^{-1}
\right\}
\bs{Y},\\
\bs{\alpha} &=
\bs{\Sigma}^{-1}_{\alpha \mid \mu}
\bs{\phi}_{\alpha \mid \mu}
\bs{S}^{-1}
\bs{Y}, \\
\bs{\beta} &= \frac{\lambda_U}{\lambda_A}\bs{\nu},
\end{align*}
where 
$\bs{\phi}_{\mu \mid\alpha}
=\bs{\phi}_\mu^\prime-\bs{\Sigma}_{\mu\alpha}\bs{\Sigma}_{\alpha\alpha}^{-1}\bs{\phi}_\alpha^\prime$,
$\bs{\phi}_{\alpha \mid\mu}
=\bs{\phi}_\alpha^\prime-\bs{\Sigma}_{\alpha\mu}\bs{\Sigma}_{\mu\mu}^{-1}\bs{\phi}_\mu^\prime$,
$\bs{\Sigma}_{\mu \mid \alpha} 
= \bs{\Sigma}_{\mu\mu} - \bs{\Sigma}_{\mu\alpha} \bs{\Sigma}_{\alpha\alpha}^{-1}\bs{\Sigma}_{\alpha\mu} $,
$\bs{\Sigma}_{\alpha \mid \mu} 
= \bs{\Sigma}_{\alpha\alpha} - \bs{\Sigma}_{\alpha\mu} \bs{\Sigma}_{\mu\mu}^{-1}\bs{\Sigma}_{\mu\alpha} $,
$\bs{\Sigma}_{\mu\mu}=\bs{\phi}_\mu^\prime\bs{S}^{-1}\bs{\phi}_\mu$,
$\bs{\Sigma}_{\mu\alpha}=\bs{\phi}_\mu^\prime\bs{S}^{-1}\bs{\phi}_\alpha$,
$\bs{\Sigma}_{\alpha\mu}=\bs{\phi}_\alpha^\prime\bs{S}^{-1}\bs{\phi}_\mu$,
$\bs{\Sigma}_{\alpha\alpha}=\bs{\phi}_\alpha^\prime\bs{S}^{-1}\bs{\phi}_\alpha$ and
$\bs{S}=\bs{M}_{\tu}+\frac{\lambda_U}{\lambda_A}\bs{R}_{\ta}=
\bs{R}_{\tu}+J\lambda_U\bs{I}+\frac{\lambda_U}{\lambda_A}\bs{R}_{\ta}$.
\end{cor}

\begin{cor}
\label{cor:lin_smoother}
Let $\bs{B}_{\mu}=\bs{\Sigma}^{-1}_{\mu \mid \alpha}
\bs{\phi}_{\mu \mid \alpha}
\bs{S}^{-1}$, 
$\bs{B}_{\nu}=\bs{S}^{-1}
\left\{
\bs{I}
-
\left(
\bs{\phi}_{\mu}\bs{\Sigma}^{-1}_{\mu \mid \alpha}\bs{\phi}_{\mu \mid \alpha}
+
\bs{\phi}_{\alpha}\bs{\Sigma}^{-1}_{\alpha \mid \mu}\bs{\phi}_{\alpha \mid \mu}
\right)
\bs{S}^{-1}
\right\}$,
$\bs{B}_{\alpha}=\bs{\Sigma}^{-1}_{\alpha \mid \mu}
\bs{\phi}_{\alpha \mid \mu}
\bs{S}^{-1}
$ 
and
$\bs{B}_{\beta}=\frac{\lambda_U}{\lambda_A}\bs{B}_{\nu}
$. The nested smoothing spline $\hat{U}(t)$ is a linear smoother, expressed in the matrix form as, $\hat{U}=\bs{K}_{\lambda_U,\lambda_A}\bs{Y}$, where $\bs{K}_{\lambda_U,\lambda_A}=\bs{\phi}_\mu\bs{B}_{\mu}+\bs{R}_{\tu}\bs{B}_{\nu}+ \bs{\phi}_\alpha\bs{B}_{\alpha}+\bs{R}_{\ta}\bs{B}_{\beta}$. 
\end{cor}
\noindent The proof is straightforward by applying Theorem \ref{thm:nSS} and Corollary \ref{cor:nSS_coef}. As a linear smoother, nSS estimates the mean function by a linear combination of observations with the weight matrix $\bs{K}_{\lambda_U,\lambda_A}$.
 
Theorem \ref{thm:nGPequalnSS} below shows the main result of this subsection, i.e. the posterior mean of U under the nGP prior is equivalent to the nSS $\hat{U}$ when $\sigma^2_{\mu} \to \infty$ and $\sigma^2_{\alpha} \to \infty$. The proof is in Appendix \ref{sec:appendix_a} and is based on the following results of Lemma \ref{lem:ECov}. 
\begin{lem}
\label{lem:ECov}
For the observations $\bs{Y}=\{Y(t_1),Y(t_2),\dots,Y(t_J) \}^\prime$ and the nested Gaussian process $U(t)$, we have
\begin{align*}
\textsf{E}\left\{ U(t) \right\}&=0,\\
\textsf{E}\left\{ \bs{Y} \right\}&=\bs{0},\\
\textsf{Cov}\left\{U(t),\bs{Y}\right\} &=\sigma_\mu^2\bs{\phi}_\mu^\prime(t)\bs{\phi}_\mu^\prime+\sigma^2_U\bs{R}_{\tu}^\prime(t)+
\sigma_\alpha^2\bs{\phi}_\alpha^\prime(t)\bs{\phi}_\alpha^\prime+\sigma^2_A\bs{R}_{\ta}^\prime(t),\\
\textsf{Cov}\left\{\bs{Y},\bs{Y}\right\} &=\sigma_\mu^2\bs{\phi}_\mu\bs{\phi}_\mu^\prime+\sigma^2_U\bs{R}_{\tu}+
\sigma_\alpha^2\bs{\phi}_\alpha\bs{\phi}_\alpha^\prime+\sigma^2_A\bs{R}_{\ta}+\sigma^2_\varepsilon\bs{I}.
\end{align*} 
\end{lem}

\begin{thm}
\label{thm:nGPequalnSS}
Given observations $\bs{Y}=\{Y(t_1),Y(t_2),\dots,Y(t_J) \}^\prime$, the posterior mean of $U(t)$ with nested Gaussian process prior is denoted as $\bar{U}_{\sigma^2_{\mu},\sigma^2_{\alpha}}(t)=\textsf{E}\{U(t) \mid \bs{Y},\sigma^2_{\mu}, \sigma^2_{\alpha}, \sigma^2_{\varepsilon} \}$. We have
\begin{equation*}
\lim \limits_{\sigma^2_{\mu} \to \infty} \lim \limits_{\sigma^2_{\alpha} \to \infty}\bar{U}_{\sigma^2_{\mu},\sigma^2_{\alpha}}(t) = \hat{U}(t), 
\end{equation*} 
where $\hat{U}(t)$ is the nested smoothing spline.
\end{thm}

\section{Posterior Computation}
To complete a Bayesian specification, we choose priors for the initial values, covariance parameters in the nGP and residual variance.  In particular, we let
$\bs{\mu} \sim \textsf{N}_m(\bs{0}, \sigma^2_{\mu}\bs{I})$, $\bs{\alpha} \sim \textsf{N}_m(\bs{0}, \sigma^2_{\alpha}\bs{I})$, $\sigma^2_\varepsilon \sim \textsf{invGamma}(a,b)$, $\sigma^2_U \sim \textsf{invGamma}(a,b)$ and $\sigma^2_U \sim \textsf{invGamma}(a,b)$, where $\textsf{invGamma}(a,b)$ denotes the inverse gamma distribution with shape parameter $a$ and scale parameter $b$. In the applications shown below, the data are rescaled so that the absolute value of the maximum observation is less than 100.  We choose diffuse but proper priors by letting $\sigma_{\mu}^{-1} = \sigma_{\alpha}^{-1} = a = b = 0.01$ as a default to allow the data to inform strongly, and have observed good performance in a variety of settings for this choice. In practice, we have found the posterior distributions for these hyperparameters to be substantially more concentrated than the prior in applications we have considered, suggesting substantial Bayesian learning.

With this prior specification, we propose an MCMC algorithm for posterior computation.  This algorithm consists of two iterative steps:
(1) Given the $\sigma^2_\varepsilon$, $\sigma^2_U$, $\sigma^2_A$ and $\bs{Y}$, draw posterior samples of $\bs{\mu}$, $\bs{U}=\{U(t_1), U(t_2), \dots, U(t_J)\}^\prime$, $\bs{\alpha}$ and $\bs{A}=\{A(t_1), A(t_2), \dots, A(t_J)\}^\prime$;
 (2) Given the $\bs{\mu}$,  $\bs{U}$, $\bs{\alpha}$,  $\bs{A}$ and $\bs{Y}$, draw posterior samples of $\sigma^2_\varepsilon$, $\sigma^2_U$ and $\sigma^2_A$.

In the first step, it would seem natural to draw ${\bf U}$ and ${\bf A}$ from their multivariate normal conditional posterior distributions.  However, this is extremely expensive computationally in high dimensions involving $O(J^3)$ computations in inverting $J \times J$ covariance matrices, which do not have any sparsity structure that can be exploited.  To reduce this computational bottleneck in GP models, there is a rich literature relying on low rank matrix approximations \citep{smola2001sparse,lawrence2002fast,quinonero2005unifying}.  Of course, such low rank approximations introduce some associated approximation error, with the magnitude of this error unknown but potentially substantial in our motivating mass spectrometry applications, as it is not clear that typical approximations having sufficiently low rank to be computationally feasible can be accurate.  

To bypass the need for such approximations, we propose a different approach that does not require inverting $J \times J$ covariance matrices but instead exploits the Markovian property implied by SDEs \eqref{eq:sde_u} and \eqref{eq:sde_a}. The Markovian property is represented by a stochastic difference equation, namely the state equation, as illustrated for the case when $m=2$ and $n=1$ in Proposition \ref{prop:se_exact} which is easily extended to cases with higher order of $m$ and $n$.              
\label{sec:comp}
\begin{prop}
\label{prop:se_exact}
When $m=2$ and $n=1$, nested Gaussian process $U(t)$ along with its first order derivative $D^1U(t)$ and $A(t)$ follow the state equation:
\begin{equation*}
\bs{\theta}_{j+1}=\bs{G}_j\bs{\theta}_{j}+\bs{\omega}_j,
\end{equation*}
where $\bs{\theta}_{j+1}=\{U(t_{j+1}), D^1U(t_{j+1}), A(t_{j+1})\}^\prime$, $\bs{\omega}_{j}  \sim \textsf{N}_3\left(\bs{0}, \bs{W}_j\right)$,
$
\bs{G}_j =
\left( {\begin{array}{ccc}
1  & \delta_j & \frac{\delta_j^2}{2} \\
0  & 1 & \delta_j \\
0  & 0 & 1 \\
\end{array} } \right)
$ and
$
\bs{W}_j =
\left( {\begin{array}{ccc}
\frac{\delta_j^3}{3}\sigma_U^2+\frac{\delta_j^5}{20}\sigma_A^2  & 
\frac{\delta_j^2}{2}\sigma_U^2+\frac{\delta_j^4}{8}\sigma_A^2 &
\frac{\delta_j^3}{6}\sigma_A^2 \\
\frac{\delta_j^2}{2}\sigma_U^2+\frac{\delta_j^4}{8}\sigma_A^2  &
\delta_j\sigma_U^2+\frac{\delta_j^3}{3}\sigma_A^2 & 
\frac{\delta_j^2}{2}\sigma_A^2 \\
\frac{\delta_j^3}{6}\sigma_A^2  & \frac{\delta_j^2}{2}\sigma_A^2  & \delta_j\sigma_A^2 \\
\end{array} } \right)
$
with $\delta_j=t_{j+1}-t_{j}$. 
\end{prop}
\noindent The proof is in Appendix \ref{sec:appendix_a}. The state equation combined with the observation equation \eqref{eq:obs} forms a state space model \citep{west1997bayesian,durbin2001time}, for which the latent states $\bs{\theta}_j$'s can be efficiently sampled by a simulation smoother algorithm \citep{durbin2002simple} with $O(J)$ computation complexity.

Given the $\bs{\mu}$, $\bs{U}$, $\bs{\alpha}$ and $\bs{A}$, posterior samples of $\sigma^2_\varepsilon$ can be obtained by drawing from the inverse-gamma conditional posterior while $\sigma_U^2$ and $\sigma_A^2$ can be updated in Metropolis-Hastings (MH) steps.  We have found that typical MH random walk steps tend to be sticky and it is preferable to use MH independence chain proposals in which one samples candidates for $\sigma_U^2$ and $\sigma_A^2$ from approximations to their conditional posteriors that are easy to sample from.  To accomplish this, we rely on the following proposition.      

\begin{prop}
\label{prop:se_approx}
When $\delta_j$ is sufficient small, the state equation in Proposition \ref{prop:se_exact} can be approximated by
\begin{equation*}
\bs{\theta}_{j+1}=\bs{\tilde{G}}_j\bs{\theta}_{j}+\bs{\tilde{H}}_j\bs{\tilde{\omega}}_j,
\end{equation*}
where $\bs{\tilde{\omega}}_{j}  \sim 
\textsf{N}_2\left(\bs{0}, \bs{\tilde{W}}_j\right)$,
$
\bs{\tilde{G}}_j =
\left( {\begin{array}{ccc}
1  & \delta_j & 0 \\
0  & 1 & \delta_j \\
0  & 0 & 1 \\
\end{array} } \right)$,
$\bs{\tilde{H}}_j =
\left( {\begin{array}{ccc}
0  & 0\\
1 & 0  \\
0  &  1  \\
\end{array} } \right)
$ and
$\bs{\tilde{W}}_j =
\left( {\begin{array}{ccc}
\sigma_U^2\delta_j  & 0  \\
0  &  \sigma_A^2\delta_j  \\
\end{array} } \right).
$
\end{prop}
\noindent The above approximate state equation is derived by applying the Euler approximation \citep[chapter 9, ][]{Kloeden92}, essentially a first-order Taylor approximation, to the SDEs \eqref{eq:sde_u} and \eqref{eq:sde_a}. Given the $\bs{\theta}_j$'s,  the $\sigma^2_U$ and $\sigma^2_A$ in the above approximate state equation can be easily sampled as a Bayesian linear regression model with the given coefficients. 

Finally, we outline the proposed MCMC algorithm as follows:
\begin{enumerate}[]
\item (1). For the state space model with the observation equation \eqref{eq:obs} and the state equation in Proposition \ref{prop:se_exact},  update the latent states $\bs{\mu}$,  $\bs{U}$, $\bs{\alpha}$ and  $\bs{A}$ by using the simulation smoother.
\item (2). Sample $\sigma^2_\varepsilon$ from the posterior distribution $\textsf{invGamma}\left(a+\frac{1}{2}J,b+\frac{1}{2}\sum_{j=1}^J\left\{Y(t_j)-U(t_j)\right\}^2\right)$.
\item (3a). Given $\sigma^2_\varepsilon$, $\sigma^2_U$ and $\sigma^2_A$, we sample the latent states $\bs{\mu}^*$,  $\bs{U}^*$, $\bs{\alpha}^*$ and $\bs{A}^*$ for the approximate state space model with the observation equation \eqref{eq:obs} and the approximate state equation specified in Proposition \ref{prop:se_approx}. 
\item (3b). Given $\bs{\mu}^*$,  $\bs{U}^*$, $\bs{\alpha}^*$ and $\bs{A}^*$,  the proposal $\sigma^{2\;*}_U$ and $\sigma^{2\;*}_A$ is drawn from the posterior distributions  $\textsf{invGamma}\left(a+\frac{1}{2}J,b+\frac{1}{2}\sum_{j=0}^{J-1}\frac{\left\{DU^*(t_{j+1})-DU^*(t_j)-A^*(t_j)\delta_j\right\}^2}{\delta_j}\right)$ and\\
$\textsf{invGamma}\left(a+\frac{1}{2}J,b+\frac{1}{2}\sum_{j=0}^{J-1}\frac{\left\{A^*(t_{j+1})-A^*(t_j)\right\}^2}{\delta_j}\right)$, respectively.
\item (3c). The proposal $\sigma^{2\;*}_U$ and $\sigma^{2\;*}_A$ will be accepted with the probability
\begin{equation*}\min \left\{
\prod_{j=0}^{J-1}\frac
{f_{\mathsf{N},3}\left(\bs{\theta}_{j+1}-\bs{G}_j\bs{\theta}_j \mid \bs{0}, \bs{W}_j^*\right)
f_{\mathsf{N},2}\left(\bs{\tilde{H}}_j(\bs{\theta}_{j+1}^*-\bs{\tilde{G}}_j\bs{\theta}_j^*) \mid \bs{0}, \bs{\tilde{W}}_j\right)}
{f_{\mathsf{N},3}\left(\bs{\theta}_{j+1}-\bs{G}_j\bs{\theta}_j \mid \bs{0}, \bs{W}_j\right)
f_{\mathsf{N},2}\left(\bs{\tilde{H}}_j(\bs{\theta}_{j+1}^*-\bs{\tilde{G}}_j\bs{\theta}_j^*) \mid \bs{0}, \bs{\tilde{W}}_j^*\right)}, 1
\right\},
\end{equation*}
where $f_{\mathsf{N},k}(\bs{X} \mid \bs{0}, \bs{\Sigma})$ denotes the probability density function of the $k$-dimensional normal random vector with mean $\bs{0}$ and covariance matrix  $\bs{\Sigma}$; $\bs{\theta}_j$, $\bs{W}_j$ and  $\bs{\tilde{W}}_j$ are specified in Proposition \ref{prop:se_exact} and \ref{prop:se_approx}; Similar notions hold for $\bs{\theta}_j^*$, $\bs{W}_j^*$ and  $\bs{\tilde{W}}_j^*$ with $\bs{\mu}$,  $\bs{U}$, $\bs{\alpha}$ $\bs{A}$, $\sigma^{2}_U$ and $\sigma^{2}_A$ replaced by $\bs{\mu}^*$,  $\bs{U}^*$, $\bs{\alpha}^*$ $\bs{A}^*$, $\sigma^{2\;*}_U$ and $\sigma^{2\;*}_A$ correspondingly.      
\end{enumerate} 

\section{Simulations}
\label{sec:simu}
We conducted a simulation study to assess the performance of the proposed method, Bayesian nonparametric regression via an nGP prior (BNR-nGP), and compared it to several alternative methods: cubic smoothing spline \citep[SS, ][]{wahba1990spline},  wavelet method with the soft minimax threshold \citep[Wavelet1, ][]{donoho1994ideal}, wavelet method with the soft Stein's unbiased estimate of risk for threshold choice \citep[Wavelet2, ][]{donoho1995adapting} and hybrid adaptive splines \citep[HAS, ][]{luo1997hybrid}. For BNR-nGP, we take the posterior mean as the estimate, which is based on the draws from the proposed MCMC algorithm with 1,500 iterations, discarding the first 500 as the burn-in stage and saving remaining ones. The other methods are implemented in R \citep{R2011}, along with the corresponding R packages for Wavelet methods \citep[wmtsa, ][]{wmtsa2010} and hybrid adaptive splines  \citep[bsml, ][]{bsml2011}.  

\begin{figure}
\centering
\subfigure[Blocks]{\includegraphics[width=0.48\textwidth]{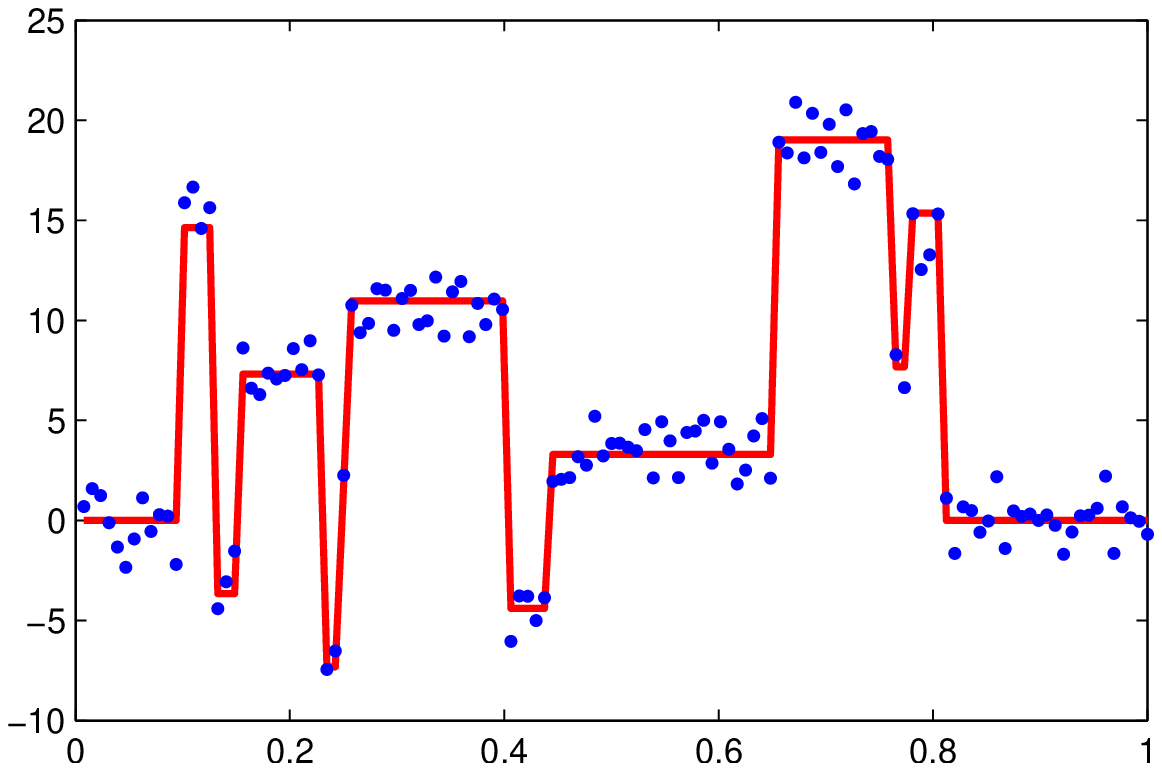}}
\subfigure[Bumps]{\includegraphics[width=0.48\textwidth]{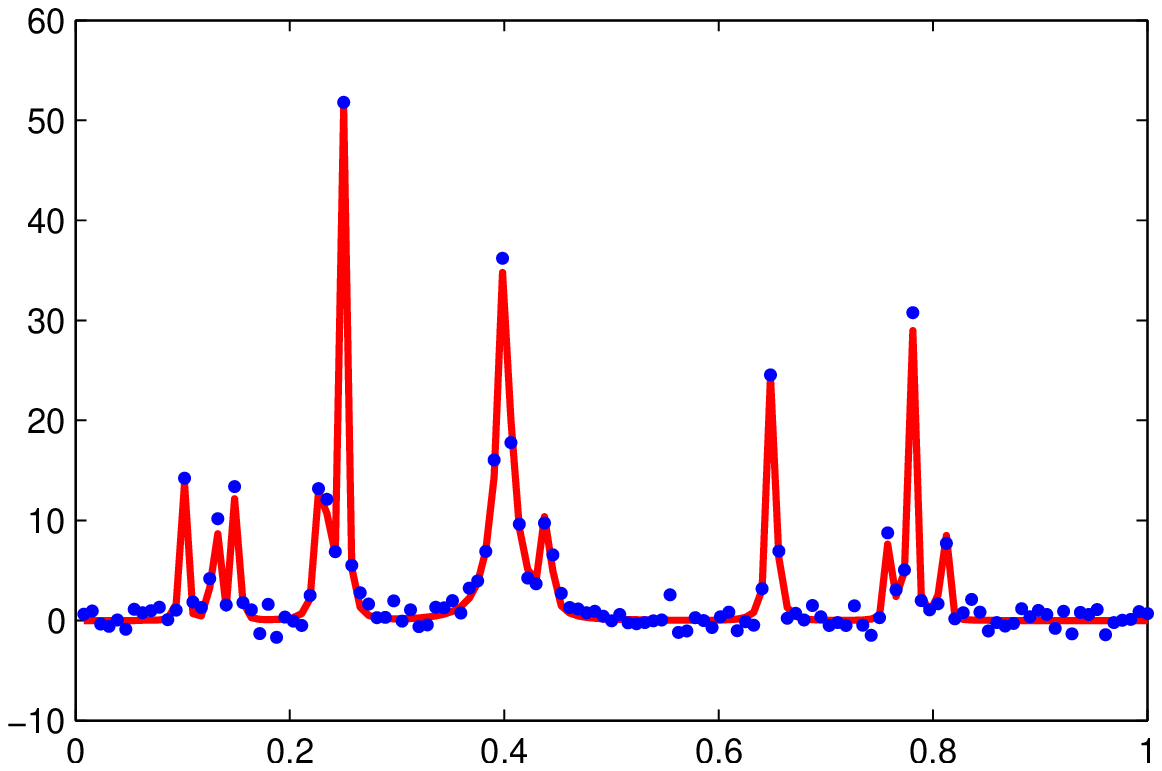}}\\
\subfigure[Heavisine]{\includegraphics[width=0.48\textwidth]{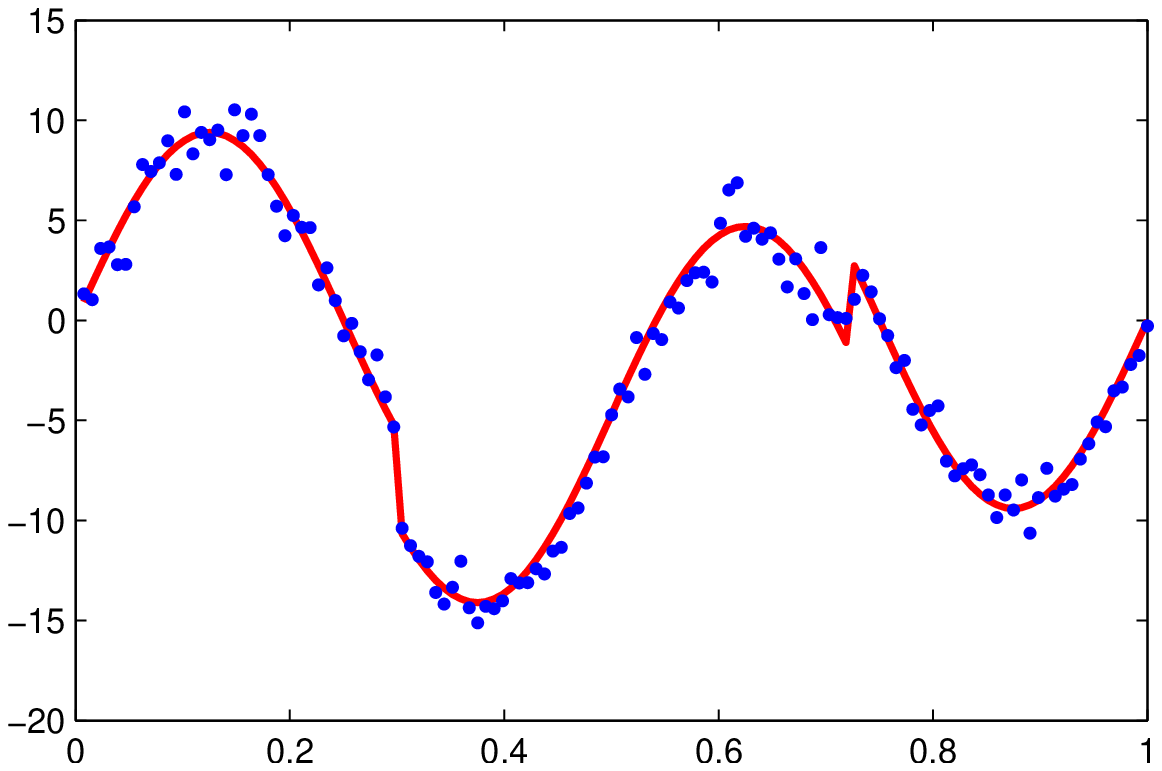}}
\subfigure[Doppler]{\includegraphics[width=0.48\textwidth]{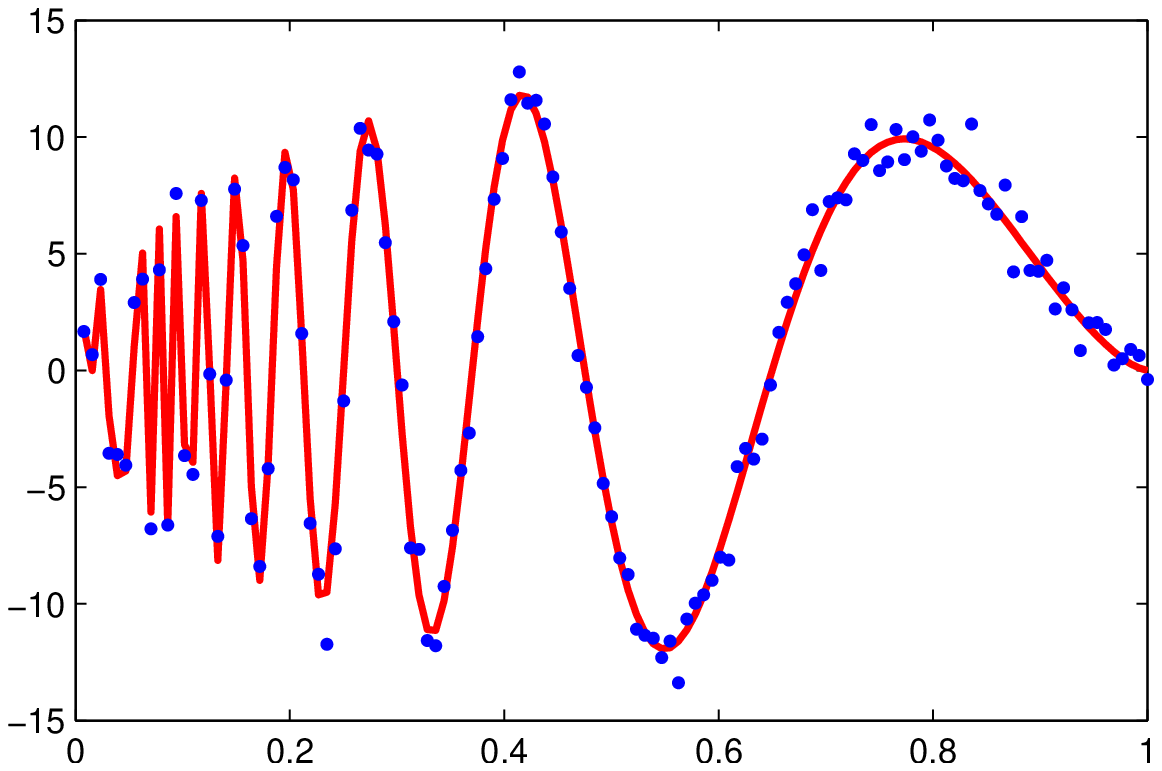}}
\caption{\label{fig:simu1} Four locally-varying smoothness functions: true function ({\color{red} \textemdash}) and 128 observations ({\color{blue} $\bullet$}).}
\end{figure}

Our first simulation study focuses on four functions adapted from \citet{donoho1994ideal}  with different types of locally-varying smoothness. The functions are plotted in Figure \ref{fig:simu1}, for which the smoothness levels vary, for example, abruptly in panel (a) or gradually in panel (d). For each function, equally-spaced observations are obtained with Gaussian noise, for which the signal-to-noise ratio is $\frac{SD(U)}{\sigma_{\varepsilon}}=7$. We use the mean squared error (MSE) $\frac{1}{J}\sum_{j=1}^J\{\hat{U}(t_j)-U_0(t_j)\}^2$ to compare the performance of different methods based on 100 replicates. The simulation results are summarized in Table \ref{tbl:simu1}. Among all methods, SS performs worst, which is not surprising since it can not adapt to the locally-varying smoothness. Among the remaining methods, BNR-nGP performs well in general for all cases with either the smallest or the second smallest average MSE across 100 replicates. In contrast, Wavelet2 and HAS may perform better for a given function, but their performances are obviously inferior for another function (e.g. Heavisine for Wavelet2 and Doppler for HAS). This suggests the nGP prior is able to adapt to a wide variety of locally-varying smoothness profiles.

\begin{figure}
\centering
\includegraphics[width=0.78\textwidth]{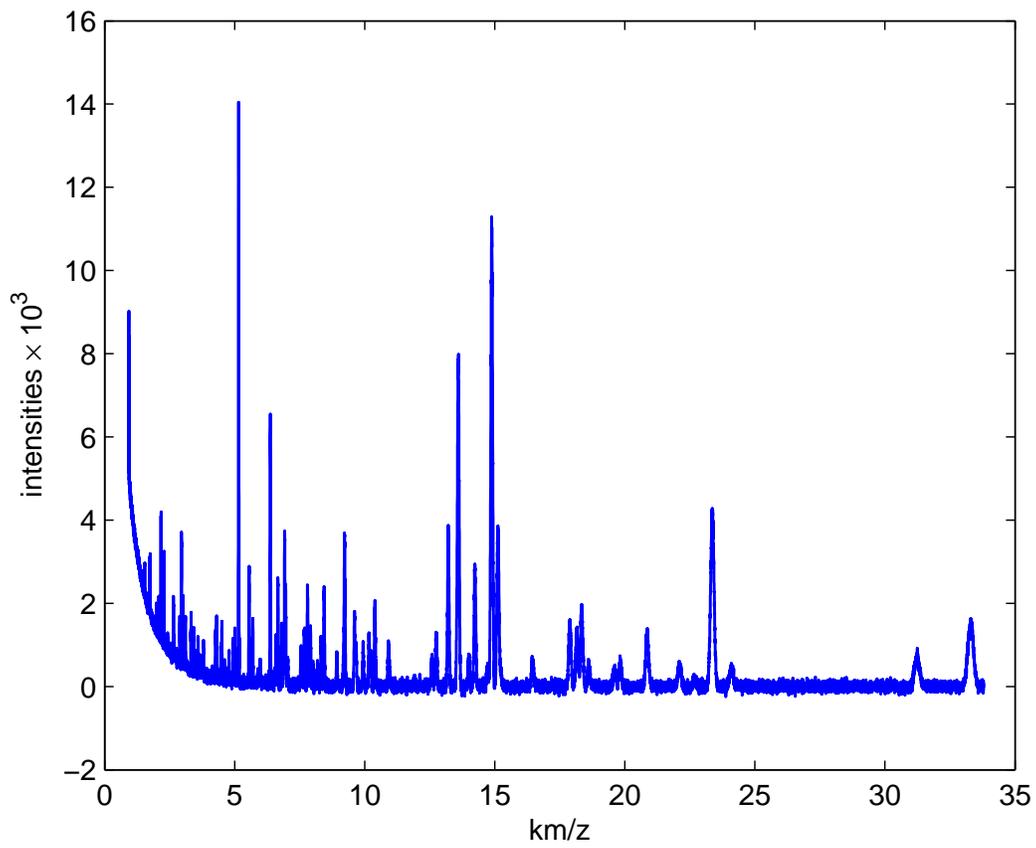}
\caption{The plot of one simulated mass spectrometry data (J=20,695).
\label{fig:PMS_simu}}
\end{figure}

We further compare the proposed method and the alternative methods for analyzing  mass spectrometry data. The 100 datasets are generated by the `virtual mass spectrometer' \citep{coombes2005understanding}, which considers the physical principles of the instrument. One set of these simulated data is plotted in Figure \ref{fig:PMS_simu} with $\sigma_\varepsilon=66$. The simulated data have been shown to accurately mimic real data  \citep{morris2005feature} and are available at \url{http://bioinformatics.mdanderson.org/Supplements/Datasets/Simulations/index.html}.
Since the analysis of all observations (J=20,695) of a given dataset is computational infeasible for HAS, we focus on the analysis of the observations within two regions with $5<km/z<8$ (region 1 with J=2,524) and $20<km/z<25$ (region 2 with J=2,235) respectively. Those two regions represent the unique feature of mass spectrometry data. More specifically, with smaller km/z values the peaks are much taller and sharper than the peaks in the region with larger km/z values.  The results in Table \ref {tbl:simu1} indicate that the BNR-nGP performs better than the other smoothness adaptive methods for both regions in terms of smaller average MSE and narrower interquartile range of MSE. Although the smoothing spline seems to work well with smaller average MSE in region 2, the peaks are clearly over-smoothed, leading to large MSEs at these important locations. In contrast, BNR-nGP had excellent performance relative to the competitors across locations. 

\begin{table}[ht]
\caption{Average MSE and the interquartile range of MSE (in parentheses) for Bayesian nonparametric regression with nGP prior (BNR-nGP), smoothing spline (SS),  wavelet method with the soft minimax threshold (Wavelet1), wavelet method with the soft Stein's unbiased estimate of risk for threshold choice (Wavelet2) and Hybrid adaptive spline (HAS).\label{tbl:simu1}}
\begin{center}
\begin{tabular}{lccccc}
 \hline\hline
Example & BNR-nGP & SS & Wavelet1 & Wavelet2  & HAS \\
\hline
Blocks    & 0.950(0.166) & 3.018(0.248) & 2.750(0.748) & 1.237(0.341)   & 0.539(0.113) \\
Bumps     & 1.014(0.185) & 26.185(0.787) & 3.433(0.938) & 1.195(0.282)  & 0.904(0.258)\\ 
Heavisine & 0.320(0.058) & 0.337(0.087) & 0.702(0.230) & 1.620(0.460)   & 0.818(0.122)\\ 
Doppler   & 0.989(0.183) & 3.403(0.361) & 1.517(0.402) & 0.695(0.179)   & 3.700(0.534)\\
MS Region 1($\times 10^{-3}$)& 1.498(0.266)&  2.293(0.513) &  2.367(0.616) & 6.048(3.441)& 72.565(39.596)\\
MS Region 2($\times 10^{-3}$)& 0.840(0.375)&  0.798(0.490) &  0.948(0.587) & 1.885(0.493)& 7.958(5.559)\\  
\hline 
\end{tabular}
\end{center}
\end{table}

\section{Applications}
\label{sec:app}
We apply the proposed method to protein mass spectrometry (MS) data.  Protein MS plays an important role in proteomics for identifying disease-related proteins in the samples \citep{cottrell1999probability,tibshirani2004sample,domon2006mass,morris2008bayesian}. For example,
Panel (a) of Figure \ref{fig:MCMC}  plots $11,186$ intensities in a pooled sample of nipple aspirate fluid from healthy breasts and breasts with cancer versus the mass to charge ratio m/z of ions \citep{coombes2005improved}. Analysis of protein MS data involves several steps, including spectra alignment, signal extraction, baseline subtraction, normalization and peak detection.  As an illustration of our method, we focus on the second step, i.e., estimate the intensity function adjusted for measurement errors.  Peaks in the intensity function may correspond to proteins that differ in the expression levels between cancer and control patients.

We fit the Bayes nonparametric regression with nGP prior and ran the MCMC algorithm for 11,000 iterations with the first 1000 iterations discarded as burn-in and every 10th draw retained for analysis. The trace plots and autocorrelation plots suggested the algorithm converged fast and mixed well. Panel (b) of Figure \ref{fig:MCMC} plots the posterior mean of $U$ and its pointwise $95\%$ credible interval. Note that the posterior mean of $U$ is adapted to the various smoothness at different regions, which is more apparently illustrated by the Panel (c) of Figure \ref{fig:MCMC}. Panel (d) of Figure \ref{fig:MCMC} demonstrates the posterior mean and $95\%$ credible interval of rate of intensity change $DU$, which suggests a peak around 4 km/z.

\section{Discussion}
We have proposed a novel nested Gaussian process prior, which is designed for flexible nonparametric locally adaptive smoothing while facilitating efficient computation even in large data sets.  Most approaches for Bayesian locally adaptive smoothing, such as free knot splines and kernel regression with varying bandwidths, encounter substantial problems with scalability.  Even isotropic Gaussian processes, which provide a widely used and studied prior for nonparametric regression, face well known issues in large data sets, with standard approaches for speeding up computation relying on low rank approximations.  It is typically not possible to assess the accuracy of such approximations and whether a low rank assumption is warranted for a particular data set.  However, when the function of interest is not smooth but can have very many local bumps and features, high resolution data may be intrinsically needed to obtain an accurate estimate of local features of the function, with low rank approximations having poor accuracy.  This seems to be the case in mass spectroscopy applications, such as the motivating proteomics example we considered in Section \ref{sec:app}.  We have simultaneously addressed two fundamental limitations of typical isotropic Gaussian process priors for nonparametric Bayes regression: (i) the lack of spatially-varying smoothness; and (ii) the lack of scalability to large sample sizes.  In addition, this was accomplished in a single coherent Bayesian probability model that fully accounts for uncertainty in the function without relying on multistage estimation.  

Although we have provided an initial study of some basic theoretical properties, the fundamental motivation in this paper is to obtain a practically useful method.  We hope that this initial work stimulates additional research along several interesting lines.  The first relates to generalizing the models and computational algorithms to multivariate regression surfaces.  Seemingly this will be straightforward to accomplish using additive models and tensor product specifications.  The second is to allow for the incorporation of prior knowledge regarding the shapes of the functions; in some applications, there is information available in the form of differential equations or even a rough knowledge of the types of curves one anticipates, which could ideally be incorporated into an nGP prior.  Finally, there are several interesting theoretical directions, such as showing rates of posterior contraction for true functions belonging to a spatially-varying smoothness class.

\label{sec:dis}

\appendix
\section{Appendix: Proofs of Theoretical Results}
\label{sec:appendix_a}
\subsection{Proof of Lemma \ref{lem:GPs_nGP}}
We specify $U(t)=\tilde{U}(t)+\tilde{A}(t)$ and $A(t)=D^m\tilde{A}(t)$. By SDEs \eqref{eq:sde_u} and \eqref{eq:sde_a},
\begin{align}
\label{eq:sde_tilde_u}
D^{m} \tilde{U}(t) &= \sigma_{U} \dot{W}_U(t),
\\
\label{eq:sde_tilde_a}
D^{m+n} \tilde{A}(t) &= \sigma_{A} \dot{W}_A(t).
\end{align}
By applying stochastic integration to SDEs \eqref{eq:sde_tilde_u} and \eqref{eq:sde_tilde_a}, it can be shown that
\begin{align*}
\tilde{U}(t)&=\tilde{U}_0(t)+\tilde{U}_1(t)=\sum_{i=0}^{m-1}\mu_i\phi_i(t)+\sigma_{U}^2\int_{\mathcal{T}}G_m(t,u)\dot{W}_U(u)du,\\
\tilde{A}(t)&=\tilde{A}_0(t)+\tilde{A}_1(t)=\sum_{i=0}^{n-1}\alpha_i\phi_{m+i}(t)+\sigma_{A}^2\int_{\mathcal{T}}G_{m+n}(t,u)\dot{W}_A(u)du
\end{align*}
given the initial values $\bs{\mu}$ and $\bs{\alpha}$. 
Since $\tilde{U}_0(t)$, $\tilde{U}_1(t)$, $\tilde{A}_0(t)$ and $\tilde{A}_1(t)$  
are the linear combination of Gaussian random variables at every $t$, they are Gaussian processes defined over $t$, whose mean functions and covariance functions can be easily derived as required. In addition, $\tilde{U}_0(t)$, $\tilde{U}_1(t)$, $\tilde{A}_0(t)$ and $\tilde{A}_1(t)$ are mutually independent due to the mutually independent assumption of $\bs{\mu}$, $\bs{\alpha}$, $\dot{W}_U(\cdot)$ and $\dot{W}_A(\cdot)$ in the definition of nGP.    

\subsection{Proof of Theorem \ref{thm:RKHS_nGP}}
We aim to characterize $\mathcal{H}_{\mathcal{K}_U}$,  the RKHS of U with the reproducing kernel $\mathcal{K}_{U}(s,t)$. The support of $U$, a mean-zero Gaussian random element, is the closure of $\mathcal{H}_{\mathcal{K}_U}$ \citep[][Lemma 5.1]{van2008reproducing}.

By Lo\`eve's Theorem \citep[][Theorem 35]{berlinet2004reproducing}, the RKHSs generated by the processes $\tilde{U}_0(t)$, $\tilde{U}_1(t)$, $\tilde{A}_0(t)$ and $\tilde{A}_1(t)$ with covariance functions $\mathcal{K}_{\tilde{U}_0}(s,t)$, $\mathcal{K}_{\tilde{U}_1}(s,t)$, $\mathcal{K}_{\tilde{A}_0}(s,t)$ and $\mathcal{K}_{\tilde{A}_1}(s,t)$ (given in Lemma \ref{lem:GPs_nGP}) are congruent to RKHSs
$\mathcal{H}_{\mathcal{K}_{\tilde{U}_0}}$,
$\mathcal{H}_{\mathcal{K}_{\tilde{U}_1}}$,
$\mathcal{H}_{\mathcal{K}_{\tilde{A}_0}}$ and 
$\mathcal{H}_{\mathcal{K}_{\tilde{A}_1}}$, respectively. Based on Theorem 5 of \citet{berlinet2004reproducing}, we conclude that $\mathcal{K}_{U}(s,t)=\mathcal{K}_{\tilde{U}_0}(s,t)+
\mathcal{K}_{\tilde{U}_1}(s,t)+
\mathcal{K}_{\tilde{A}_0}(s,t)+
\mathcal{K}_{\tilde{A}_1}(s,t)$ is the reproducing kernel of the RKHS \begin{align*}
\mathcal{H}_{\mathcal{K}_U}=&\;\mathcal{H}_{\mathcal{K}_{\tilde{U}_0}}\oplus\mathcal{H}_{\mathcal{K}_{\tilde{U}_1}}\oplus
\mathcal{H}_{\mathcal{K}_{\tilde{A}_0}}\oplus\mathcal{H}_{\mathcal{K}_{\tilde{A}_1}}\\
=&\;\{U(t): U(t)=\tilde{U}_0(t)+\tilde{U}_1(t)+\tilde{A}_0(t)+\tilde{A}_1(t),\\ &\;\hspace{38pt}\tilde{U}_0(t) \in \mathcal{H}_{\mathcal{K}_{\tilde{U}_0}},
\tilde{U}_1(t) \in \mathcal{H}_{\mathcal{K}_{\tilde{U}_1}},
\tilde{A}_0(t) \in \mathcal{H}_{\mathcal{K}_{\tilde{A}_0}},
\tilde{A}_1(t) \in \mathcal{H}_{\mathcal{K}_{\tilde{A}_1}} \}
\end{align*}

\subsection{Proof of Theorem \ref{thm:pos_con}}
Similar to the proof of Lemma \ref{lem:GPs_nGP}, we specify 
$U=\tilde{U}_0+\tilde{U}_1+\tilde{A}_0+\tilde{A}_1$, which is a mean zero Gaussian process with continuous and differentiable covariance function $\mathcal{K}_{U}(s,t)=\mathcal{K}_{\tilde{U}_0}(s,t)+\mathcal{K}_{\tilde{U}_1}(s,t)+\mathcal{K}_{\tilde{A}_0}(s,t)+\mathcal{K}_{\tilde{A}_1}(s,t)$. 


We aims to verify the sufficient conditions of the strong consistency theorem \citep[Theorem 1, ][]{choi2007posterior} for nonparametric regression: (I) prior positivity of neighborhoods and (II) existence of uniformly exponentially consistent tests and sieves $\Theta_J$ with $\Pi_U(\Theta_J^C) \leq C_1\exp(-C_2J)$ for some positive constants $C_1$ and $C_2$.

Given $U$ is a Gaussian process with continuous sample path and continuous covariance function, it follows from Theorem 4 of \citet{ghosal2006posterior} that $\Pi_U( ||U - U_0 ||_\infty < \delta) > 0$ for any $\delta > 0$. In addition, for every $\delta >0$, $\Pi_{\sigma_\varepsilon}\left(|\frac{\sigma_\varepsilon}{\sigma_{\varepsilon, 0}}-1| <\delta\right)>0$ under Assumption \ref{assum:posprior}. Hence, we can define a neighborhood $B_\delta=\left\{ (U, \sigma_\varepsilon): ||U - U_0 ||_\infty < \delta, |\frac{\sigma_\varepsilon}{\sigma_{\varepsilon, 0}}-1| <\delta\right\}$ such that $\Pi_{(U,  \sigma_\varepsilon)}(B_\delta)>0$ satisfying the condition (I). 

From Theorem 2 of \citet{choi2007posterior}, we can show that for a sequence of $M_J$,  there exist uniformly exponentially consistent tests for the sieves $\Theta_J=\left\{U: ||U||_\infty < M_J, ||DU||_\infty < M_J\right\}$ under the infill design Assumption \ref{assum:design}. What remains is to verify the exponentially small probability of $\Theta_{J,0}^C=\left\{U: ||U||_\infty > M_J\right\}$ and $\Theta_{J,1}^C=\left\{U: ||DU||_\infty > M_J\right\}$. Using Borell's inequality \citep[Proposition A.2.7, ][]{van1996weak}, we have 
\begin{equation*}
\Pi_U\left(||U||_\infty > M_J\right) \leq C_1\exp\left(-\frac{C_3M_J^2}{\sigma^2}\right)
\end{equation*}
for some positive constants $C_1$ and $C_3$,  and
\begin{align*}
\sigma^2:=\;&\sup_{t\in[0,t_U]} \left\{ E(\tilde{U}_0+\tilde{U}_1+\tilde{A}_0+\tilde{A}_1)^2 \right\}\\
=\;& \sup_{t\in[0,t_U]} \left\{ E\tilde{U}_0^2+E\tilde{U}_1^2+E\tilde{A}_0^2+E\tilde{A}_1^2 \right\}\\
=\;& \sigma^2_{\mu}\sum_{i=0}^{m-1}\phi_i^2(t_U)+\frac{\sigma_U^2t_U^{2m-1}}{(m-1)!(m-1)!(2m-1)}+ \\
\;& \sigma^2_{\alpha}\sum_{i=0}^{n-1}\phi_{m+i}^2(t_U)+\frac{\sigma_A^2t_U^{2m+2n-1}}{(m+n-1)!(m+n-1)!(2m+2n-1)}.  
\end{align*}
By applying the Borel\myhyph Cantelli theorem, we have 
\begin{equation*}
\Pi_U\left(||U||_\infty > M_J\right) \leq C_1\exp\left(-C_2J\right),
\end{equation*}
almost surely under the exponential tail Assumption \ref{assum:exp}. By the similar arguments, we can show that $\Pi_U\left(||DU||_\infty > M_J\right) \leq C_1\exp(-C_2J)$. 

Hence, the conditions (I) and (II) hold, which leads to the strong consistency for Bayesian nonparametric regression with nGP prior.

\subsection{Proof of Corollary \ref{cor:GP_pss}}
Note that $\tilde{U}(t)=\tilde{U}_0(t)+\tilde{U}_1(t)=\sum_{i=0}^{m-1}\mu_i\phi_i(t)+\sigma_{U}^2\int_{\mathcal{T}}G_m(t,u)\dot{W}_U(u)du
$ is the prior for the polynomial smoothing spline \citep[][Section 1.5]{wahba1990spline}.   By the similar arguments in Theorem \ref{thm:RKHS_nGP}, we can show that the support of $\tilde{U}$  is the closure of RKHS $\mathcal{H}_{\mathcal{K}_{\tilde{U}}}=\mathcal{H}_{\mathcal{K}_{\tilde{U}_0}}\oplus\mathcal{H}_{\mathcal{K}_{\tilde{U}_1}}$. 

Thus, $\mathcal{K}_{U}(s,t)-\mathcal{K}_{\tilde{U}}(s,t)=\mathcal{K}_{\tilde{A}}(s,t)=\mathcal{H}_{\mathcal{K}_{\tilde{A}_0}}\oplus\mathcal{H}_{\mathcal{K}_{\tilde{A}_1}}$  a nonnegative kernel, which implies that $\mathcal{H}_{\mathcal{K}_{\tilde{U}}} \subset \mathcal{H}_{\mathcal{K}_{U}}$ by Corollary 4 of \citet{aronszajn1950theory}. 

\subsection{Proof of Theorem \ref{thm:nSS}}
Let $U(t)=\tilde{U}(t)+\tilde{A}(t)$ and $A(t)=D^m\tilde{A}(t)$. The nested penalized sum-of-square \eqref{eq:npss} can be written as:
\begin{equation}
\label{eq:nPSS1}
\textsf{nPSS}(t)=
\frac{1}{J}\sum_{j=1}^J\left\{Y(t_j)-\tilde{U}(t_j)-\tilde{A}(t_j)\right\}^2 + \lambda_{U}\int_{\mathcal{T}}\left\{D^{m}\tilde{U}(t)\right\}^2dt +
\lambda_{A}\int_{\mathcal{T}}\left\{D^{m+n}\tilde{A}(t)\right\}^2dt,
\end{equation}
where $\tilde{U}(t)$ is the $m$-order polynomial smoothing spline and $\tilde{A}(t)$ is the $(m+n)$-order polynomial smoothing spline. 

By the classical RKHS theory of the polynomial smoothing spline \citep[Section 1.2]{wahba1990spline}, there exists a unique decomposition of $\tilde{U}(t)$: 
\begin{align*}
\tilde{U}(t)&=\tilde{U}_0(t)+\tilde{U}_1(t)\\
&=\sum_{i=0}^{m-1}\mu_i\phi_i(t)+\int_{\mathcal{T}} G_m(t,u) D^m \tilde{U}(t) d u
\end{align*}
with $\tilde{U}_0(t) \in \mathcal{H}_{\mathcal{R}_{\tilde{U}_0}}$ and $\tilde{U}_1(t) \in \mathcal{H}_{\mathcal{R}_{\tilde{U}_1}}$. $\mathcal{H}_{\mathcal{R}_{\tilde{U}_0}}=\left\{f(t): D^m f(t)=0, t \in \mathcal{T}\right\}$ nad \\$\mathcal{H}_{\mathcal{R}_{\tilde{U}_1}}=\left\{f(t): D^if(t)\text{ absolutely continuous for } i=0,1,\cdots,m-1, D^mf(t) \in \mathcal{L}_2(\mathcal{T}) \right\}$ are the RKHSs with reproducing kernel $\mathcal{R}_{\tilde{U}_0}(s,t)$ and $\mathcal{R}_{\tilde{U}_1}(s,t)$ respectively, where $\phi_i(t)$, $G_m(t,u)$, $\mathcal{R}_{\tilde{U}_0}(s,t)$ and $\mathcal{R}_{\tilde{U}_1}(s,t)$ are defined in Theorem \ref{thm:RKHS_nGP} with $\mathcal{L}_2(\mathcal{T})=\left\{f(t): \int_{\mathcal{T}}f^2(t)dt < \infty\right\}$ the space of squared integrable functions defined on index set $\mathcal{T}$. 

Given $\mathcal{T}_o=\left\{t_j: j=1,2,\cdots,J\right\}$, the $\tilde{U}_1(t) \in \mathcal{H}_{\mathcal{R}_{\tilde{U}_1}}$ can be uniquely written as $\tilde{U}_1(t)=\sum_{j=1}^{J}\nu_j\mathcal{R}_{\tilde{U}_1}(t_j,t)+\eta_{\tilde{U}_1}(t)$, where $\eta_{\tilde{U}_1}(\cdot) \in \mathcal{H}_{\mathcal{R}_{\tilde{U}_1}}$ orthogonal to $\mathcal{R}_{\tilde{U}_1}(t_j,\cdot)$ with inner product \\$\langle\mathcal{R}_{\tilde{U}_1}(t_j,\cdot),\eta_{\tilde{U}_1}(\cdot)\rangle_{\mathcal{H}_{\mathcal{R}_{\tilde{U}_1}}}=
\int_{\mathcal{T}}D^m\mathcal{R}_{\tilde{U}_1}(t_j,u) D^m\eta_{\tilde{U}_1}(u)du=0$ for $j=1,2,\cdots,J$. 

As a result,
\begin{align*}
\int_{\mathcal{T}}\left\{D^{m}\tilde{U}(t)\right\}^2dt &=\int_{\mathcal{T}}\left[D^{m}\left\{\sum_{i=0}^{m-1}\mu_i\phi_i(t)+
\sum_{j=1}^{J}\nu_j\mathcal{R}_{\tilde{U}_1}(t_j,t)+
\eta_{\tilde{U}_1}(t)\right\}\right]^2dt\\
&=\sum_{j=1}^{J}\sum_{j^\prime=1}^{J}\nu_j\mathcal{R}_{\tilde{U}_1}(t_j,t_{j^\prime})\nu_{j^\prime}+
\int_{\mathcal{T}}\left\{D^m\eta_{\tilde{U}_1}(t)\right\}^2dt\\
&=\bs{\nu}^\prime\bs{R}_{\tu}\bs{\nu}+
\langle\eta_{\tilde{U}_1}(\cdot),\eta_{\tilde{U}_1}(\cdot)\rangle_{\mathcal{H}_{\mathcal{R}_{\tilde{U}_1}}}.
\end{align*}

By similar arguments,
\begin{align*}
\tilde{A}(t)&=\tilde{A}_0(t)+\tilde{A}_1(t)\\
&=\sum_{i=0}^{n-1}\alpha_i\phi_{m+i}(t)+
\sum_{j=1}^{J}\beta_j\mathcal{R}_{\ta_1}(t_j,t)+\eta_{\ta_1}(t),
\end{align*}
and
\begin{align*}
\int_{\mathcal{T}}\left\{D^{m}\tilde{A}(t)\right\}^2dt =
\bs{\beta}^\prime\bs{R}_{\ta}\bs{\beta}+
\langle\eta_{\tilde{A}_1}(\cdot),\eta_{\tilde{A}_1}(\cdot)\rangle_{\mathcal{H}_{\mathcal{R}_{\ta_1}}},
\end{align*}
where $\tilde{A}_0(t) \in \mathcal{H}_{\mathcal{R}_{\tilde{A}_0}}$ and $\tilde{A}_1(t) \in \mathcal{H}_{\mathcal{R}_{\tilde{A}_1}}$ with $\mathcal{H}_{\mathcal{R}_{\tilde{A}_0}}=\left\{f(t): D^{m+n} f(t)=0, t \in \mathcal{T}\right\}$ and \\$\mathcal{H}_{\mathcal{R}_{\tilde{A}_1}}=\left\{f(t): D^if(t)\text{ absolutely continuous for } i=0,1,\cdots,m+n-1, D^{m+n}f(t) \in \mathcal{L}_2(\mathcal{T}) \right\}$ the RKHSs with reproducing kernel $\mathcal{R}_{\tilde{A}_0}(s,t)$ and $\mathcal{R}_{\tilde{A}_1}(s,t)$ respectively; 
$\eta_{\tilde{A}_1}(\cdot) \in \mathcal{H}_{\mathcal{R}_{\tilde{A}_1}}$ is orthogonal to $\mathcal{R}_{\tilde{A}_1}(t_j,\cdot)$ with inner product $\langle\mathcal{R}_{\tilde{A}_1}(t_j,\cdot),\eta_{\tilde{A}_1}(\cdot)\rangle_{\mathcal{H}_{\mathcal{R}_{\tilde{A}_1}}}=
\int_{\mathcal{T}}D^m\mathcal{R}_{\tilde{A}_1}(t_j,u) D^m\eta_{\tilde{A}_1}(u)du=0$ for $j=1,2,\cdots,J$. 

Note that $\eta_{\tilde{U}_1}(t_j)=\langle\mathcal{R}_{\tilde{U}_1}(t_j,\cdot),\eta_{\tilde{U}_1}(\cdot)\rangle_{\mathcal{H}_{\mathcal{R}_{\tilde{U}_1}}}=0$ and
$\eta_{\tilde{A}_1}(t_j)=\langle\mathcal{R}_{\tilde{A}_1}(t_j,\cdot),\eta_{\tilde{A}_1}(\cdot)\rangle_{\mathcal{H}_{\mathcal{R}_{\tilde{A}_1}}}=0$ due to the reproducing property of  $\mathcal{R}_{\tilde{U}_1}(t_j,\cdot)$ and $\mathcal{R}_{\tilde{A}_1}(t_j,\cdot)$.     
It then follows from expression \eqref{eq:nPSS1} that
\begin{align*}
\textsf{nPSS}(t)=
&\frac{1}{J}\left(\bs{Y}-\bs{\phi}_{\mu}\bs{\mu}-\bs{R}_{\tu}\bs{\nu}-\bs{\phi}_{\alpha}\bs{\alpha}-\bs{R}_{\ta}\bs{\beta}\right)^\prime
\left(\bs{Y}-\bs{\phi}_{\mu}\bs{\mu}-\bs{R}_{\tu}\bs{\nu}-\bs{\phi}_{\alpha}\bs{\alpha}-\bs{R}_{\ta}\bs{\beta}\right)\\
&+\lambda_U\bs{\nu}^\prime\bs{R}_{\tu}\bs{\nu}+\lambda_A\bs{\beta}^\prime\bs{R}_{\ta}\bs{\beta}+
\langle\eta_{\tilde{U}_1}(\cdot),\eta_{\tilde{U}_1}(\cdot)\rangle_{\mathcal{H}_{\mathcal{R}_{\tilde{U}_1}}}+
\langle\eta_{\tilde{A}_1}(\cdot),\eta_{\tilde{A}_1}(\cdot)\rangle_{\mathcal{H}_{\mathcal{R}_{\ta_1}}},
\end{align*}
which is minimized when $\langle\eta_{\tilde{U}_1}(\cdot),\eta_{\tilde{U}_1}(\cdot)\rangle_{\mathcal{H}_{\mathcal{R}_{\tilde{U}_1}}}=
\langle\eta_{\tilde{A}_1}(\cdot),\eta_{\tilde{A}_1}(\cdot)\rangle_{\mathcal{H}_{\mathcal{R}_{\ta_1}}}=0$. Thus, $\eta_{\tilde{U}_1}(\cdot)=\eta_{\tilde{A}_1}(\cdot)=0$ and we obtain the forms of $\hat{U}(t)$ and $\textsf{nPSS(t)}$ as required. 

\subsection{Proof of Corollary \ref{cor:nSS_coef}}
We first take partial derivatives of nested penalized sum-of-squares $\textsf{nPSS}(t)$ in Theorem \ref{thm:nSS} with respective to $\bs{\mu}$, $\bs{\nu}$, $\bs{\alpha}$ and $\bs{\beta}$ and set them to zeros:
\begin{align}
\label{eq:d_mu}
\frac{\partial\;\textsf{nPSS}(t)}{\partial\;\bs{\mu}}&=
\bs{\phi}_\mu^\prime\left(\bs{\phi}_{\mu}\bs{\mu}+\bs{R}_{\tu}\bs{\nu}+\bs{\phi}_{\alpha}\bs{\alpha}+\bs{R}_{\ta}\bs{\beta}-\bs{Y}\right)=\bs{0},\\
\label{eq:d_nu}
\frac{\partial\;\textsf{nPSS}(t)}{\partial\;\bs{\nu}}&=
\bs{R}_{\tu}\left(\bs{\phi}_{\mu}\bs{\mu}+\bs{M}_{\tu}\bs{\nu}+\bs{\phi}_{\alpha}\bs{\alpha}+\bs{R}_{\ta}\bs{\beta}-\bs{Y}\right)=\bs{0},\\
\label{eq:d_alpha}
\frac{\partial\;\textsf{nPSS}(t)}{\partial\;\bs{\alpha}}&=
\bs{\phi}_\alpha^\prime\left(\bs{\phi}_{\mu}\bs{\mu}+\bs{R}_{\tu}\bs{\nu}+\bs{\phi}_{\alpha}\bs{\alpha}+\bs{R}_{\ta}\bs{\beta}-\bs{Y}\right)=\bs{0},\\
\label{eq:d_beta}
\frac{\partial\;\textsf{nPSS}(t)}{\partial\;\bs{\beta}}&=
\bs{R}_{\ta}\left(\bs{\phi}_{\mu}\bs{\mu}+\bs{R}_{\tu}\bs{\nu}+\bs{\phi}_{\alpha}\bs{\alpha}+\bs{M}_{\ta}\bs{\beta}-\bs{Y}\right)=\bs{0},
\end{align}
where $\bs{M}_{\tu}=\bs{R}_{\tu}+J\lambda_U\bs{I}$ and $\bs{M}_{\ta}=\bs{R}_{\ta}+J\lambda_A\bs{I}$.
It follows from equations \eqref{eq:d_nu} and \eqref{eq:d_beta} that 
\begin{align*}
\bs{\nu}&=\bs{S}^{-1}\left(\bs{Y}-\bs{\phi}_{\mu}\bs{\mu}-\bs{\phi}_{\alpha}\bs{\alpha}\right),\\
\bs{\beta}&=\frac{\lambda_U}{\lambda_A}\bs{\nu}. 
\end{align*}
Substituting them into equations \eqref{eq:d_mu} and \eqref{eq:d_alpha} with some algebra leads to
\begin{align*}
\bs{\Sigma}_{\mu\mu}\bs{\mu}+\bs{\Sigma}_{\mu\alpha}\bs{\alpha}
&=\bs{\phi}_\mu^\prime \bs{S}^{-1}\bs{Y},\\
\bs{\Sigma}_{\alpha\mu}\bs{\mu}+\bs{\Sigma}_{\alpha\alpha}\bs{\alpha}
&=\bs{\phi}_\alpha^\prime \bs{S}^{-1}\bs{Y},
\end{align*}
from which we obtain
\begin{align*}
\bs{\mu}&=
\left(
\bs{\Sigma}_{\mu\mu} - \bs{\Sigma}_{\mu\alpha} \bs{\Sigma}_{\alpha\alpha}^{-1}\bs{\Sigma}_{\alpha\mu}
\right)^{-1}
\left(
\bs{\phi}_\mu^\prime-\bs{\Sigma}_{\mu\alpha}\bs{\Sigma}_{\alpha\alpha}^{-1}\bs{\phi}_\alpha^\prime
\right)
\bs{S}^{-1}\bs{Y}
=
\bs{\Sigma}_{\mu \mid \alpha}^{-1}\bs{\phi}_{\mu \mid\alpha}\bs{S}^{-1}\bs{Y},\\
\bs{\alpha}&=
\left(
\bs{\Sigma}_{\alpha\alpha} - \bs{\Sigma}_{\alpha\mu} \bs{\Sigma}_{\mu\mu}^{-1}\bs{\Sigma}_{\mu\alpha}
\right)^{-1}
\left(
\bs{\phi}_\alpha^\prime-\bs{\Sigma}_{\alpha\mu}\bs{\Sigma}_{\mu\mu}^{-1}\bs{\phi}_\mu^\prime
\right)
\bs{S}^{-1}\bs{Y}
=
\bs{\Sigma}_{\alpha \mid \mu}^{-1}\bs{\phi}_{\alpha \mid \mu}\bs{S}^{-1}\bs{Y}.   
\end{align*}
It is then straightforward to show
\begin{align*}
\bs{\nu} &= \bs{S}^{-1}
\left\{
\bs{I}
-
\left(
\bs{\phi}_{\mu}\bs{\Sigma}^{-1}_{\mu \mid \alpha}\bs{\phi}_{\mu \mid \alpha}
+
\bs{\phi}_{\alpha}\bs{\Sigma}^{-1}_{\alpha \mid \mu}\bs{\phi}_{\alpha \mid \mu}
\right)
\bs{S}^{-1}
\right\}
\bs{Y},\\
\bs{\beta} &= \frac{\lambda_U}{\lambda_A}\bs{\nu}
\end{align*}
as desired.

\subsection{Proof of Lemma \ref{lem:ECov}}
Let $U(t)=\tilde{U}(t)+\tilde{A}(t)$ and $A(t)=D^m\tilde{A}(t)$. From SDEs \eqref{eq:sde_u} and \eqref{eq:sde_a},
\begin{align*}
D^{m} \tilde{U}(t) &= \sigma_{U} \dot{W}_U(t),
\\
D^{m+n} \tilde{A}(t) &= \sigma_{A} \dot{W}_A(t).
\end{align*}
Thus, given the initial value $\bs{\mu}$, it can be shown that $\tilde{U}(t)=\sum_{i=0}^{m-1}\mu_i\phi_i(t)+\sigma_{U}^2\int_{\mathcal{T}}G_m(t,u)\dot{W}_U(u)du$, a $(m-1)$-fold integrated Wiener process \citep{shepp1966radon}. Similarly,
$\tilde{A}(t)=\sum_{i=0}^{n-1}\alpha_i\phi_{m+i}(t)+\sigma_{A}^2\int_{\mathcal{T}}G_{m+n}(t,u)\dot{W}_A(u)du$, a $(m+n-1)$-fold integrated Wiener process.
 
It is obvious that $\textsf{E}\left\{ U(t) \right\}=0$ and 
$\textsf{E}\left\{ \bs{Y} \right\}=\bs{0}$. Given the mutually independent assumption of $\bs{\mu}$, $\bs{\alpha}$, $\dot{W}_U(\cdot)$ and $\dot{W}_A(\cdot)$,
\begin{align*}
\textsf{Cov}\left\{U(t), Y(t_j)\right\} =\;&\textsf{Cov}\left\{U(t), U(t_j)\right\} \\
=\; &\textsf{Cov}\left\{\tilde{U}(t), \tilde{U}(t_j)\right\}+
\textsf{Cov}\left\{\tilde{A}(t), \tilde{A}(t_j)\right\}\\
=\;&\textsf{E}\left\{\tilde{U}(t) \tilde{U}(t_j)\right\}+
\textsf{E}\left\{\tilde{A}(t) \tilde{A}(t_j)\right\}\\
=\;&\sigma_{\mu}^2 \sum_{i=0}^{m-1}\phi_i(t)\phi_i(t_j)+
\sigma_{U}^2 \mathcal{R}_{\tilde{U}_1}(t,t_j)+\\
&\sigma_{\alpha}^2 \sum_{i=0}^{n-1}\phi_{m+i}(t)\phi_{m+i}(t_j)+
\sigma_{A}^2 \mathcal{R}_{\tilde{A}_1}(t,t_j),
\end{align*}
and
\begin{align*}
\textsf{Cov}\left\{Y(t_j), Y(t_{j^\prime})\right\} =\;&\textsf{Cov}\left\{U(t_j), U(t_{j^\prime})\right\}+
\sigma_{\varepsilon}^2\\
=\;&\sigma_{\mu}^2 \sum_{i=0}^{m-1}\phi_i(t_{j})\phi_i(t_{j^\prime})+
\sigma_{U}^2 \mathcal{R}_{\tilde{U}_1}(t_j,t_{j^\prime})+\\
&\sigma_{\alpha}^2 \sum_{i=0}^{n-1}\phi_{m+i}(t_j)\phi_{m+i}(t_{j^\prime})+
\sigma_{A}^2 \mathcal{R}_{\tilde{A}_1}(t_j,t_{j^\prime})+
\sigma_{\varepsilon}^2,
\end{align*}  
for $j=1,2,\cdots,J$ and $j^\prime=1,2,\cdots,J$. The lemma holds.

\subsection{Proof of Theorem \ref{thm:nGPequalnSS}}
By Lemma \ref{lem:ECov} and the  results on conditional multivariate normal distribution \citep{searle1982matrix},
\begin{align*}
\textsf{E}\{U(t) \mid \bs{Y},\sigma^2_{\mu}, \sigma^2_{\alpha}, \sigma^2_{\varepsilon} \} =\;& \textsf{Cov}\left\{U(t),\bs{Y}\right\}
\textsf{Cov}^{-1}\left\{\bs{Y},\bs{Y}\right\}\bs{Y}\\
=\;&\left[
\rho_\mu\bs{\phi}_\mu^\prime(t)\bs{\phi}_\mu^\prime+\bs{R}_{\tu}^\prime(t)+
\rho_\alpha\bs{\phi}_\alpha^\prime(t)\bs{\phi}_\alpha^\prime+\rho_A\bs{R}_{\ta}^\prime(t)
\right] \times\\
&\left[\
\rho_\mu\bs{\phi}_\mu\bs{\phi}_\mu^\prime+\rho_\alpha\bs{\phi}_\alpha\bs{\phi}_\alpha^\prime+
\rho_A\bs{R}_{\ta}+\bs{R}_{\tu}
+J\lambda_U\bs{I}
\right]^{-1}\bs{Y}\\
=\;&\bs{\phi}_\mu^\prime(t)\left(\rho_\mu\bs{\phi}_\mu^\prime\bs{\Sigma}_{\rho_\mu\rho_\alpha}^{-1}\right)\bs{Y}+
\bs{R}_{\tu}^\prime(t)\bs{\Sigma}_{\rho_\mu\rho_\alpha}^{-1}\bs{Y}+\\
&\bs{\phi}_\alpha^\prime(t)\left(\rho_\alpha\bs{\phi}_\alpha^\prime\bs{\Sigma}_{\rho_\mu\rho_\alpha}^{-1}\right)\bs{Y}+
\rho_A\bs{R}_{\ta}^\prime(t)\bs{\Sigma}_{\rho_\mu\rho_\alpha}^{-1}\bs{Y}
\end{align*}
where $\rho_\mu=\sigma^2_\mu/\sigma^2_U$, $\rho_\alpha=\sigma^2_\alpha/\sigma^2_U$,
$\rho_A=\sigma^2_A/\sigma^2_U$, $J\lambda_U=\sigma^2_\varepsilon/\sigma^2_U$ and $\bs{\Sigma}_{\rho_\mu\rho_\alpha}=\rho_\mu\bs{\phi}_\mu\bs{\phi}_\mu^\prime+\bs{S}_{\rho_\alpha}$ with $\bs{S}_{\rho_\alpha}=\rho_\alpha\bs{\phi}_\alpha\bs{\phi}_\alpha^\prime+\bs{S}=\rho_\alpha\bs{\phi}_\alpha\bs{\phi}_\alpha^\prime+\rho_A\bs{R}_{\ta}+\bs{R}_{\tu}
+J\lambda_U\bs{I}$. We are going to evaluate the limits of $\rho_\mu\bs{\phi}_\mu^\prime\bs{\Sigma}_{\rho_\mu\rho_\alpha}^{-1}$, $\rho_\alpha\bs{\phi}_\alpha^\prime\bs{\Sigma}_{\rho_\mu\rho_\alpha}^{-1}$ and $\bs{\Sigma}_{\rho_\mu\rho_\alpha}^{-1}$ when $\rho_\mu \to +\infty$ and  $\rho_\alpha \to +\infty$.
  
It can be verified  that
\begin{align}
\label{eq:sigma_limit}
\bs{\Sigma}_{\rho_\mu\rho_\alpha}^{-1}
&=
\bs{S}_{\rho_\alpha}^{-1}-
\bs{S}_{\rho_\alpha}^{-1}\bs{\phi}_\mu
\left(\bs{\phi}_\mu^\prime\bs{S}_{\rho_\alpha}^{-1}\bs{\phi}_\mu\right)^{-1}
\left\{\bs{I}+\rho_\mu^{-1}\left(\bs{\phi}_\mu^\prime\bs{S}_{\rho_\alpha}^{-1}\bs{\phi}_\mu\right)^{-1}\right\}^{-1}\bs{\phi}_\mu^\prime\bs{S}_{\rho_\alpha}^{-1},\\
\bs{S}_{\rho_\alpha}^{-1}
&=
\bs{S}^{-1}-
\bs{S}^{-1}\bs{\phi}_\alpha
\left(\bs{\phi}_\alpha^\prime\bs{S}^{-1}\bs{\phi}_\alpha\right)^{-1}
\left\{\bs{I}+\rho_\alpha^{-1}\left(\bs{\phi}_\alpha^\prime\bs{S}^{-1}\bs{\phi}_\alpha\right)^{-1}\right\}^{-1}\bs{\phi}_\alpha^\prime\bs{S}^{-1}. \notag
\end{align}
It follows that $\bs{S}_{\infty}^{-1}=\lim\limits_{\rho_\alpha \to +\infty}\bs{S}_{\rho_\alpha}^{-1}=\bs{S}^{-1}-
\bs{S}^{-1}\bs{\phi}_\alpha
\left(\bs{\phi}_\alpha^\prime\bs{S}^{-1}\bs{\phi}_\alpha\right)^{-1}
\bs{\phi}_\alpha^\prime\bs{S}^{-1}=\bs{S}^{-1}-
\bs{S}^{-1}\bs{\phi}_\alpha
\bs{\Sigma}_{\alpha\alpha}^{-1}
\bs{\phi}_\alpha^\prime\bs{S}^{-1}$ and  $\bs{\phi}_\mu^\prime\bs{S}_{\infty}^{-1}\bs{\phi}_\mu=\bs{\Sigma}_{\mu \mid \alpha}$ and $\bs{S}_{\infty}^{-1}\bs{\phi}_\mu=\bs{S}^{-1}\bs{\phi}_{\mu \mid \alpha}^\prime$. 

As a result,
\begin{align*}
\bs{\Sigma}_{\infty \infty}^{-1}
&=
\lim\limits_{\rho_\mu \to +\infty} \lim\limits_{\rho_\alpha \to +\infty}\bs{\Sigma}_{\rho_\mu\rho_\alpha}^{-1}\\
&=\bs{S}^{-1}-\bs{S}^{-1}\left(\bs{\phi}_\alpha\bs{\Sigma}_{\alpha\alpha}^{-1}\bs{\phi}_\alpha^\prime+\bs{\phi}_{\mu \mid \alpha}^\prime\bs{\Sigma}_{\mu \mid \alpha}^{-1}\bs{\phi}_{\mu \mid \alpha}\right)\bs{S}^{-1}\\
&=\bs{S}^{-1}-\bs{S}^{-1}
\left\{
\bs{\phi}_\mu\bs{\Sigma}_{\mu \mid \alpha}^{-1}\bs{\phi}_{\mu \mid \alpha}+
\bs{\phi}_{\alpha}
\left(
\bs{\Sigma}_{\alpha \alpha}^{-1}\bs{\phi}_{\alpha}^\prime-
\bs{\Sigma}_{\alpha \alpha}^{-1}\bs{\Sigma}_{\alpha \mu}\bs{\Sigma}_{\mu \mid \alpha}^{-1}\bs{\phi}_{\mu \mid \alpha}
\right)
\right\}\bs{S}^{-1}\\
&=\bs{S}^{-1}\left\{\bs{I}-
\left(
\bs{\phi}_\mu\bs{\Sigma}_{\mu \mid \alpha}^{-1}\bs{\phi}_{\mu \mid \alpha}+
\bs{\phi}_{\alpha}\bs{\Sigma}_{\alpha \mid \mu}^{-1}\bs{\phi}_{\alpha \mid \mu}
\right)\bs{S}^{-1}
\right\}.
\end{align*}

By expression \eqref{eq:sigma_limit},
\begin{align*}
\rho_\mu\bs{\phi}_\mu^\prime\bs{\Sigma}_{\rho_\mu\rho_\alpha}^{-1} 
&=
\rho_\mu
\left[
\bs{I}-
\left\{
\bs{I}+\rho_\mu^{-1}
\left(
\bs{\phi}_\mu^\prime\bs{S}_{\rho_\alpha}^{-1}\bs{\phi}_\mu
\right)^{-1}
\right\}^{-1}
\right]
\bs{\phi}_\mu^\prime\bs{S}_{\rho_\alpha}^{-1}\\
&=\left(
\bs{\phi}_\mu^\prime\bs{S}_{\rho_\alpha}^{-1}\bs{\phi}_\mu
\right)^{-1}
\left\{
\bs{I}+\rho_\mu^{-1}
\left(
\bs{\phi}_\mu^\prime\bs{S}_{\rho_\alpha}^{-1}\bs{\phi}_\mu
\right)^{-1}
\right\}^{-1}
\bs{\phi}_\mu^\prime\bs{S}_{\rho_\alpha}^{-1}.
\end{align*}
It follows that $\lim\limits_{\rho_\mu \to +\infty} \lim\limits_{\rho_\alpha \to +\infty}\rho_\mu\bs{\phi}_\mu^\prime\bs{\Sigma}_{\rho_\mu\rho_\alpha}^{-1}
=\bs{\Sigma}_{\mu \mid \alpha}^{-1}\bs{\phi}_{\mu \mid \alpha}\bs{S}^{-1}$. By similar arguments, $\lim\limits_{\rho_\mu \to +\infty} \lim\limits_{\rho_\alpha \to +\infty}\rho_\alpha\bs{\phi}_\alpha^\prime\bs{\Sigma}_{\rho_\mu\rho_\alpha}^{-1}=\bs{\Sigma}_{ \alpha \mid \mu}^{-1}\bs{\phi}_{\alpha \mid \mu}\bs{S}^{-1}$.

Hence, $\bs{\Sigma}_{\infty \infty}^{-1}\bs{Y}=\bs{\nu}$, $\rho_A\bs{\Sigma}_{\infty \infty}^{-1}\bs{Y}=\bs{\beta}$,
$\lim\limits_{\rho_\mu \to +\infty} \lim\limits_{\rho_\alpha \to +\infty}\rho_\mu\bs{\phi}_\mu^\prime\bs{\Sigma}_{\rho_\mu\rho_\alpha}^{-1}\bs{Y}=\bs{\mu}$  and  
$\lim\limits_{\rho_\mu \to +\infty} \lim\limits_{\rho_\alpha \to +\infty}\rho_\alpha\bs{\phi}_\alpha^\prime\bs{\Sigma}_{\rho_\mu\rho_\alpha}^{-1}\bs{Y}=\bs{\alpha}$. The theorem holds.

\subsection{Proof of Proposition \ref{prop:se_exact}}
When $m=2$ and $n=1$, the SDEs \eqref{eq:sde_u} and \eqref{eq:sde_a} can be written as,
\begin{align*}
D^1 \bs{\theta}(t)=\bs{C}\bs{\theta}(t)+\bs{D}\bs{\dot{W}}(t),
\end{align*}
where $\bs{\theta}(t)=\left\{\begin{array}{c} U(t)  \\ D^1U(t) \\ A(t)
\end{array}  \right\}$, 
$\bs{C}=
\left( {\begin{array}{ccc}
0 & 1 & 0 \\
0 & 0 & 1 \\
0 & 0 & 0 
\end{array} } \right),
$
$\bs{C}=
\left( {\begin{array}{cc}
0 & 0 \\
\sigma_U & 0  \\
0 & \sigma_A  
\end{array} } \right)
$ and
$
\bs{\dot{W}}(t)=\left\{{\begin{array}{c} \dot{W}_U(t) \\\dot{W}_A(t) 
\end{array} } \right\}.
$

As a result,
\begin{align*}
\bs{\theta}_{j+1}&=\exp(\bs{C}\delta_j)\bs{\theta}_j+
\int_0^{\delta_j}\exp\{\bs{C}(\delta_j-u)\}\bs{D}\bs{\dot{W}}(t_j+u)du\\
&=\bs{G}_j\bs{\theta}_j+\bs{\omega}_j,
\end{align*}
where $\bs{G}_j=\exp(\bs{C}\delta_j)=\bs{I}+\delta_j\bs{C}+\delta_j^2\bs{C}\bs{C}/{2}=\left( {\begin{array}{ccc}
1  & \delta_j & \frac{\delta_j^2}{2} \\
0  & 1 & \delta_j \\
0  & 0 & 1 \\
\end{array} } \right)$ and $\bs{\omega}_{j}  \sim \textsf{N}_3\left(\bs{0}, \bs{W}_j\right)$ with
\begin{align*}
\bs{W}_j&=\int_0^{\delta_j}\exp\{\bs{C}(\delta_j-u)\}\bs{D}\bs{D}^\prime\exp\{\bs{C}^\prime(\delta_j-u)\}du\\
&=
\left( {\begin{array}{ccc}
\frac{\delta_j^3}{3}\sigma_U^2+\frac{\delta_j^5}{20}\sigma_A^2  & 
\frac{\delta_j^2}{2}\sigma_U^2+\frac{\delta_j^4}{8}\sigma_A^2 &
\frac{\delta_j^3}{6}\sigma_A^2 \\
\frac{\delta_j^2}{2}\sigma_U^2+\frac{\delta_j^4}{8}\sigma_A^2  &
\delta_j\sigma_U^2+\frac{\delta_j^3}{3}\sigma_A^2 & 
\frac{\delta_j^2}{2}\sigma_A^2 \\
\frac{\delta_j^3}{6}\sigma_A^2  & \frac{\delta_j^2}{2}\sigma_A^2  & \delta_j\sigma_A^2 \\
\end{array} } \right)
\end{align*}
as required.
\bibliographystyle{asa}
\bibliography{nGP}

\begin{thebibliography}{54}
\newcommand{\enquote}[1]{``#1''}
\expandafter\ifx\csname natexlab\endcsname\relax\def\natexlab#1{#1}\fi

\bibitem[{Abramovich and Steinberg(1996)}]{abramovich1996improved}
Abramovich, F. and Steinberg, {\relax D.M}. (1996), \enquote{Improved inference
  in nonparametric regression using L-smoothing splines,} \textit{Journal of
  Statistical Planning and Inference}, 49, 327--341.

\bibitem[{Aronszajn(1950)}]{aronszajn1950theory}
Aronszajn, N. (1950), \enquote{Theory of Reproducing Kernels,}
  \textit{Transactions of the American Mathematical Society}, 68, 337--404.

\bibitem[{Berlinet and Thomas-Agnan(2004)}]{berlinet2004reproducing}
Berlinet, A. and Thomas-Agnan, C. (2004), \textit{Reproducing kernel Hilbert
  spaces in probability and statistics}, Netherlands: Springer.

\bibitem[{Bhattacharya et~al.(2011)Bhattacharya, Pati, and
  Dunson}]{Bhattacharya2011}
Bhattacharya, A., Pati, D., and Dunson, {\relax D.B}. (2011), \enquote{Adaptive
  dimension reduction with a Gaussian process prior,} \textit{Arxiv preprint
  arXiv:1111.1044}.

\bibitem[{Choi and Schervish(2007)}]{choi2007posterior}
Choi, T. and Schervish, {\relax M.J}. (2007), \enquote{On posterior consistency
  in nonparametric regression problems,} \textit{Journal of Multivariate
  Analysis}, 98, 1969--1987.

\bibitem[{Constantine and Percival(2010)}]{wmtsa2010}
Constantine, W. and Percival, D. (2010), \textit{wmtsa: Insightful Wavelet
  Methods for Time Series Analysis},
  \url{http://CRAN.R-project.org/package=wmtsa}.

\bibitem[{Coombes et~al.(2005{\natexlab{a}})Coombes, Koomen, Baggerly, Morris,
  and Kobayashi}]{coombes2005understanding}
Coombes, K., Koomen, J., Baggerly, K., Morris, J., and Kobayashi, R.
  (2005{\natexlab{a}}), \enquote{Understanding the characteristics of mass
  spectrometry data through the use of simulation,} \textit{Cancer
  Informatics}, 1, 41.

\bibitem[{Coombes et~al.(2005{\natexlab{b}})Coombes, Tsavachidis, Morris,
  Baggerly, Hung, and Kuerer}]{coombes2005improved}
Coombes, {\relax K.R.}., Tsavachidis, S., Morris, {\relax J.S}., Baggerly,
  {\relax K.A}., Hung, {\relax M.C}., and Kuerer, {\relax H.M}.
  (2005{\natexlab{b}}), \enquote{Improved peak detection and quantification of
  mass spectrometry data acquired from surface-enhanced laser desorption and
  ionization by denoising spectra with the undecimated discrete wavelet
  transform,} \textit{Proteomics}, 5, 4107--4117.

\bibitem[{Cottrell and London(1999)}]{cottrell1999probability}
Cottrell, J. and London, U. (1999), \enquote{Probability-based protein
  identification by searching sequence databases using mass spectrometry data,}
  \textit{Electrophoresis}, 20, 3551--3567.

\bibitem[{Crainiceanu et~al.(2007)Crainiceanu, Ruppert, Carroll, Joshi, and
  Goodner}]{crainiceanu2007spatially}
Crainiceanu, {\relax C.M}., Ruppert, D., Carroll, {\relax R.J}., Joshi, A., and
  Goodner, B. (2007), \enquote{Spatially adaptive Bayesian penalized splines
  with heteroscedastic errors,} \textit{Journal of Computational and Graphical
  Statistics}, 16, 265--288.

\bibitem[{Denison et~al.(1998)Denison, Mallick, and
  Smith}]{denison1998automatic}
Denison, {\relax D.G.T}., Mallick, {\relax B.K}., and Smith, {\relax A.F.M}.
  (1998), \enquote{Automatic Bayesian curve fitting,} \textit{Journal of the
  Royal Statistical Society: Series B (Statistical Methodology)}, 60, 333--350.

\bibitem[{Dimatteo et~al.(2001)Dimatteo, Genovese, and
  Kass}]{dimatteo2001bayesian}
Dimatteo, I., Genovese, {\relax C.R}., and Kass, {\relax R.E}. (2001),
  \enquote{Bayesian curve-fitting with free-knot splines,} \textit{Biometrika},
  88, 1055.

\bibitem[{Domon and Aebersold(2006)}]{domon2006mass}
Domon, B. and Aebersold, R. (2006), \enquote{Mass spectrometry and protein
  analysis,} \textit{Science}, 312, 212.

\bibitem[{Donoho and Johnstone(1995)}]{donoho1995adapting}
Donoho, {\relax D.L}. and Johnstone, {\relax I.M}. (1995), \enquote{Adapting to
  unknown smoothness via wavelet shrinkage,} \textit{Journal of the American
  Statistical Association}, 1200--1224.

\bibitem[{Donoho and Johnstone(1994)}]{donoho1994ideal}
Donoho, {\relax D.L}. and Johnstone, {\relax J.M}. (1994), \enquote{Ideal
  spatial adaptation by wavelet shrinkage,} \textit{Biometrika}, 81, 425--455.

\bibitem[{Durbin and Koopman(2002)}]{durbin2002simple}
Durbin, J. and Koopman, S. (2002), \enquote{A simple and efficient simulation
  smoother for state space time series analysis,} \textit{Biometrika}, 89, 603.

\bibitem[{Durbin and Koopman(2001)}]{durbin2001time}
Durbin, J. and Koopman, {\relax S.J}. (2001), \textit{{Time series analysis by
  state space methods}}, vol.~24, Oxford: Oxford University Press.

\bibitem[{Fan and Gijbels(1995)}]{fan1995data}
Fan, J. and Gijbels, I. (1995), \enquote{Data-driven bandwidth selection in
  local polynomial fitting: variable bandwidth and spatial adaptation,}
  \textit{Journal of the Royal Statistical Society. Series B (Statistical
  Methodology)}, 371--394.

\bibitem[{Friedman(1991)}]{friedman1991multivariate}
Friedman, {\relax J.H}. (1991), \enquote{Multivariate adaptive regression
  splines,} \textit{The Annals of Statistics}, 1--67.

\bibitem[{Friedman and Silverman(1989)}]{friedman1989flexible}
Friedman, {\relax J.H}. and Silverman, {\relax B.W}. (1989), \enquote{Flexible
  parsimonious smoothing and additive modeling,} \textit{Technometrics}, 3--21.

\bibitem[{George and McCulloch(1993)}]{george1993variable}
George, {\relax E.I}. and McCulloch, {\relax R.E}. (1993), \enquote{Variable
  selection via Gibbs sampling,} \textit{Journal of the American Statistical
  Association}, 881--889.

\bibitem[{Ghosal and Roy(2006)}]{ghosal2006posterior}
Ghosal, S. and Roy, A. (2006), \enquote{Posterior consistency of Gaussian
  process prior for nonparametric binary regression,} \textit{The Annals of
  Statistics}, 34, 2413--2429.

\bibitem[{Green(1995)}]{green1995reversible}
Green, {\relax P.J}. (1995), \enquote{Reversible jump Markov chain Monte Carlo
  computation and Bayesian model determination,} \textit{Biometrika}, 82, 711.

\bibitem[{Heckman and Ramsay(2000)}]{heckman2000penalized}
Heckman, {\relax N.E}. and Ramsay, {\relax J.O}. (2000), \enquote{Penalized
  regression with model-based penalties,} \textit{Canadian Journal of
  Statistics}, 28, 241--258.

\bibitem[{Kloeden and Platen(1992)}]{Kloeden92}
Kloeden, {\relax P.E}. and Platen, E. (1992), \textit{{Numerical Solution of
  Stochastic Differential Equations}}, New York: Springer Verlag.

\bibitem[{Lawrence et~al.(2002)Lawrence, Seeger, and
  Herbrich}]{lawrence2002fast}
Lawrence, {\relax N.D}., Seeger, M., and Herbrich, R. (2002), \enquote{Fast
  sparse Gaussian process methods: The informative vector machine,}
  \textit{Advances in neural information processing systems}, 15, 609--616.

\bibitem[{Luo and Wahba(1997)}]{luo1997hybrid}
Luo, Z. and Wahba, G. (1997), \enquote{Hybrid adaptive splines,}
  \textit{Journal of the American Statistical Association}, 107--116.

\bibitem[{Morris et~al.(2005)Morris, Coombes, Koomen, Baggerly, and
  Kobayashi}]{morris2005feature}
Morris, J., Coombes, K., Koomen, J., Baggerly, K., and Kobayashi, R. (2005),
  \enquote{Feature extraction and quantification for mass spectrometry in
  biomedical applications using the mean spectrum,} \textit{Bioinformatics},
  21, 1764--1775.

\bibitem[{Morris et~al.(2008)Morris, Brown, Baggerly, and
  Coombes}]{morris2008bayesian}
Morris, {\relax J.S}., Brown, {\relax P.J}and~Herrick, {\relax R.C}., Baggerly,
  {\relax K.A}., and Coombes, {\relax K.R}. (2008), \enquote{Bayesian Analysis
  of Mass Spectrometry Proteomic Data Using Wavelet-Based Functional Mixed
  Models,} \textit{Biometrics}, 64, 479--489.

\bibitem[{Neal(1998)}]{neal1998regression}
Neal, R. (1998), \enquote{Regression and classification using gaussian process
  priors,} \textit{Bayesian Statistics}, 6, 475--501.

\bibitem[{Pintore et~al.(2006)Pintore, Speckman, and
  Holmes}]{pintore2006spatially}
Pintore, A., Speckman, P., and Holmes, {\relax C.C}. (2006), \enquote{Spatially
  adaptive smoothing splines,} \textit{Biometrika}, 93, 113.

\bibitem[{Quinonero-Candela and Rasmussen(2005)}]{quinonero2005unifying}
Quinonero-Candela, J. and Rasmussen, {\relax C.E}. (2005), \enquote{A unifying
  view of sparse approximate Gaussian process regression,} \textit{The Journal
  of Machine Learning Research}, 6, 1939--1959.

\bibitem[{{R Development Core Team}(2011)}]{R2011}
{R Development Core Team} (2011), \textit{R: A Language and Environment for
  Statistical Computing}, R Foundation for Statistical Computing, Vienna,
  Austria, \url{http://www.R-project.org}.

\bibitem[{Rasmussen and Williams(2006)}]{rasmussen2006gaussian}
Rasmussen, {\relax C.E}. and Williams, {\relax C.K.I}. (2006), \textit{Gaussian
  processes for machine learning}, Boston: MIT Press.

\bibitem[{Ruppert and Carroll(2000)}]{ruppert2000spatially}
Ruppert, D. and Carroll, {\relax R.J}. (2000), \enquote{Spatially-adaptive
  Penalties for Spline Fitting,} \textit{Australian \& New Zealand Journal of
  Statistics}, 42, 205--223.

\bibitem[{Savitsky et~al.(2011)Savitsky, Vannucci, and
  Sha}]{savitsky2011variable}
Savitsky, T., Vannucci, M., and Sha, N. (2011), \enquote{Variable selection for
  nonparametric Gaussian process priors: Models and computational strategies,}
  \textit{Statistical Science}, 26, 130--149.

\bibitem[{Searle(1982)}]{searle1982matrix}
Searle, {\relax S.R}. (1982), \textit{Matrix Algebra Useful for Statistics},
  New York: Wiley.

\bibitem[{Shepp(1966)}]{shepp1966radon}
Shepp, {\relax L.A}. (1966), \enquote{Radon-Nikodym derivatives of Gaussian
  measures,} \textit{The Annals of Mathematical Statistics}, 37, 321--354.

\bibitem[{Shi and Choi(2011)}]{shi1011gaussian}
Shi, {\relax J.Q}. and Choi, {\relax T}. (2011), \textit{Gaussian Process
  Regression Analysis for Functional Data}, {London: Chapman \& Hall/CRC
  Press}.

\bibitem[{Smith and Kohn(1996)}]{smith1996nonparametric}
Smith, M. and Kohn, R. (1996), \enquote{Nonparametric regression using Bayesian
  variable selection,} \textit{Journal of Econometrics}, 75, 317--343.

\bibitem[{Smola and Bartlett(2001)}]{smola2001sparse}
Smola, {\relax A.J}. and Bartlett, P. (2001), \enquote{Sparse greedy Gaussian
  process regression,} in \textit{Advances in Neural Information Processing
  Systems 13}, Citeseer.

\bibitem[{Tibshirani et~al.(2004)Tibshirani, Hastie, Narasimhan, Soltys, Shi,
  Koong, and Le}]{tibshirani2004sample}
Tibshirani, R., Hastie, T., Narasimhan, B., Soltys, S., Shi, G., Koong, A., and
  Le, {\relax Q.T}. (2004), \enquote{Sample classification from protein mass
  spectrometry, by 'peak probability contrasts',} \textit{Bioinformatics}, 20,
  3034--3044.

\bibitem[{Van~der Vaart and Van~Zanten(2008{\natexlab{a}})}]{van2008rates}
Van~der Vaart, {\relax A.W}. and Van~Zanten, {\relax J.H}.
  (2008{\natexlab{a}}), \enquote{Rates of contraction of posterior
  distributions based on Gaussian process priors,} \textit{The Annals of
  Statistics}, 36, 1435--1463.

\bibitem[{Van~der Vaart and
  Van~Zanten(2008{\natexlab{b}})}]{van2008reproducing}
--- (2008{\natexlab{b}}), \enquote{Reproducing kernel Hilbert spaces of
  Gaussian priors,} \textit{Limits of Contemporary Statistics: Contributions in
  Honor of Jayanta K. Ghosh}, 3, 200--222.

\bibitem[{Van~der Vaart and Wellner(1996)}]{van1996weak}
Van~der Vaart, {\relax A.W}. and Wellner, {\relax J.A}. (1996), \textit{Weak
  convergence and empirical processes}, New York: Springer Verlag.

\bibitem[{Wahba(1990)}]{wahba1990spline}
Wahba, G. (1990), \textit{Spline models for observational data}, vol.~59,
  Philadelphia: Society for Industrial Mathematics.

\bibitem[{Wahba(1995)}]{Wahba1995adpative}
--- (1995), \enquote{Discussion of a paper by Donoho et al.} \textit{Journal of
  the Royal Statistical Society. Series B (Statistical Methodology)}, 360--361.

\bibitem[{West and Harrison(1997)}]{west1997bayesian}
West, M. and Harrison, J. (1997), \textit{{Bayesian Forecasting and Dynamic
  Models}}, New York: Springer Verlag.

\bibitem[{Wolpert et~al.(2011)Wolpert, \relax{M.A}, and
  \relax{C.}}]{Wolpert2011Stochastic}
Wolpert, {\relax R.L}., \relax{M.A}, C., and \relax{C.}, T. (2011),
  \enquote{Stochastic expansions using continuous dictionaries: L\'{e}vy
  adaptive regression kernels,} \textit{The Annals of Statistics}, 39,
  1916--1962.

\bibitem[{Wood et~al.(2002)Wood, Jiang, and Tanner}]{wood2002bayesian}
Wood, {\relax S.A}., Jiang, W., and Tanner, M. (2002), \enquote{Bayesian
  mixture of splines for spatially adaptive nonparametric regression,}
  \textit{Biometrika}, 89, 513.

\bibitem[{Wu et~al.(2011)Wu, Sklar, Wang, and Meiring}]{bsml2011}
Wu, {\relax J.Q}., Sklar, J., Wang, {\relax Y.D}., and Meiring, W. (2011),
  \textit{bsml: Basis Selection from Multiple Libraries},
  \url{http://CRAN.R-project.org/package=bsml}.

\bibitem[{Zhou and Shen(2001)}]{zhou2001spatially}
Zhou, S. and Shen, X. (2001), \enquote{Spatially adaptive regression splines
  and accurate knot selection schemes,} \textit{Journal of the American
  Statistical Association}, 96, 247--259.

\bibitem[{Zhu et~al.(2011)Zhu, Song, and Taylor}]{zhu2011stochastic}
Zhu, B., Song, {\relax P.X.K}., and Taylor, {\relax J.M.G}. (2011),
  \enquote{Stochastic Functional Data Analysis: A Diffusion Model-Based
  Approach,} \textit{Biometrics. In press}.

\bibitem[{Zou et~al.(2010)Zou, Huang, Lee, and
  Hoeschele}]{zou2010nonparametric}
Zou, F., Huang, H., Lee, S., and Hoeschele, I. (2010), \enquote{Nonparametric
  Bayesian Variable Selection With Applications to Multiple Quantitative Trait
  Loci Mapping With Epistasis and Gene--Environment Interaction,}
  \textit{Genetics}, 186, 385.

\end{thebibliography}

\end{document}